\documentclass[preprint]{rmaa}

\usepackage{paralist}

\usepackage{psfrag,color}

\title{Optical Line profile variability of the B1+neutron star binary system\\
LS {\rm I}+65 010 = 2S\,0114+650}

\author{Gloria Koenigsberger\altaffilmark{1,2}
		Gabriela Canalizo\altaffilmark{3}
		Anabel Arrieta\altaffilmark{1}
		Michael G. Richer\altaffilmark{1}
		Leonid Georgiev\altaffilmark{1}
}

\altaffiltext{1}{Instituto de Astronom\'{\i}a, UNAM, M\'exico D.F. 04510}
\altaffiltext{2}{Centro de Ciencias F\'{\i}sicas, UNAM, M\'exico}
\altaffiltext{3}{Institute of Geophysics and Planetary Sciences, Lawrence Livermore National Laboratory}

\shortauthor{Koenigsberger et al.}
\shorttitle{Optical Line profile variability...}

\fulladdresses{
\item Anabel Arrieta,Leonid Georgiev, Gloria Koenigsberger, Michael Richer: 
Instituto de Astronom\'{\i}a, UNAM, Apdo. Postal 70-264, M\'exico D.F. 04510
(anabel,georgiev,gloria,richer@astroscu.unam.mx)

\item Gloria Koenigsberger:Centro de Ciencias F\'{\i}sicas, UNAM, Apdo. Postal 48-3, 
Cuernavaca,Mor, 62251, M\'exico

\item Gabriela Canalizo:Institute of Geophysics and Planetary Sciences, Lawrence Livermore
National Laboratory, 7000 East Avenue, L-413, Livermore, CA 94550}

\listofauthors{Gloria Koenigsberger, Gabriela Canalizo, Anabel Arrieta,
			Michael G. Richer \& Leonid Georgiev}

\indexauthor{Koenigsberger, G.}
\indexauthor{Canalizo, G.}
\indexauthor{Arrieta, A.}
\indexauthor{Richer, M.G.}
\indexauthor{Georgiev,L.N.}

\abstract{

We present high resolution spectroscopic observations of the massive X-ray
binary system LS {\rm I}+65 010=2S\,0114+650 in the optical wavelength region.
A correlation between equivalent width and radial velocity of photospheric absorption 
lines is found. The systemic velocity, inferred from the weaker 
lines is v$_{helio}$ $-$31$\pm$5 km s$^{-1}$, which, if attributed solely to the 
Galactic rotation curve, implies that LS {\rm I}$+$65 010 lies within 3 kpc 
from the Sun. The ISM Na I D lines display 2 resolved high velocity components 
at v$_{helio}$ $-$70, $-$48 km s$^{-1}$, possibly associated with gas surrounding the
binary system, in addition to the $-$24 $-$8 km s$^{-1}$ ISM features due to the Orion 
and in the Perseus arm regions.  Strong photospheric line profile variability is present 
on a night to night timescale, with He I 5875 \AA\ displaying an additional 
blue-shifted absorption in some of the spectra, similar to what is observed in 
the optical counterpart of Vela X-1. A connection between the extended blue wing and 
X-ray maximum is suggested.  Short timescale variations in line profiles are detected on 
only two nights, but the evidence that these variations occur on the 2.78 hour X-ray flaring 
period is marginal.  
}

\resumen{
Presentamos los resultados de observaciones espectroscopicas de alta 
resoluci\'on del sistema binario masivo con emision de rayos-X LS {\rm I}+65 
010=2S\,0114+650.  Se encuentra una correlaci\'on entre el ancho equivalente y 
la velocidad radial de las  l\'{\i}neas en absorci\'on tal que las l\'{\i}neas 
m\'as intensas tienen las velocidades mas grandes.  La velocidad del sistema se 
deduce utilizando las l\'{\i}neas menos intensas, obteni\'endose v$_{helio}$=$-31$km 
s$^{-1}$, la cual, si se atribuye unicamente a la rotaci\'on Gal\'actica, implica que 
LS {\rm I}$+$65 010 se ubica a $\sim$3 kpc del la vecindad solar.  Las lineas 
interestelares de Na I D muestran dos componentes de alta velocidad y dos componentes 
que se pueden asociar a los brazos espirales de Orion y de Perseo.  Se reporta la 
presencia de variabilidad en los perfiles de las las l\'{\i}neas, y se sugiere que la 
curva de velocidad radial del sistema est\'a deformada por estas variaciones.  Nuestros 
datos favorecen el periodo orbital P$_{orb}$=11.591.
}

\addkeyword{massive X-ray binary systems; 2S \,0114+650}

\begin{document}
\maketitle

\section{Introduction}

The study of massive X-ray binary systems (MXRBs) is of significant
interest because it provides a wealth of information on a  variety of
astrophysical problems  such as the determination of the masses of the
collapsed objects, the mechanisms that lead to the emission of X-rays,
the physical processes involved in the production of accretion disks and
relativistic jets, and possibly the phenomena that result in some of
the observed $\gamma$-ray bursts. Progress in solving some of these
problems, however, requires precise information of
the binary system parameters. In particular, well-determined  properties
of the normal star in the system are essential. This is, however,
not an easy enterprise in most cases due to the effects introduced by 
the interactions that occur in the system.
For example, the X-rays that are produced near the collapsed companion
can alter the ionization structure of the stellar wind of the primary,
producing emission that contaminates the photospheric absorption-line
spectrum.  If the X-ray flux is variable, the absorption line profiles
may also reflect this variability.  In addition, the presence of
a close companion may produce stellar pulsations if the orbital period is not
synchronized with the rotational period (Kumar, Ao \& Quataert 1995, Witte \& 
Savonije 1999), and
such pulsations are expected to produce line-profile variability 
(Vogt \& Penrod 1983).  It is therefore of interest to establish the degree to 
which
photospheric lines are altered by the presence of the collapsed companion
and hence, detailed studies of individual binary systems are necessary.
In this paper we present an analysis of line profile variability in LS {\rm 
I}$+$65 010,
the optical counterpart of the intriguing MXRB, 2S0114$+$650.

First discovered as an X-ray source by SAS-3 in 1977 (Dower et al. 1977), the
optical counterpart of 2S0114$+$650 was identified shortly thereafter (Margon 
1977).  Its variable star name is V662 Cas (Kholopov et al., 1989).  The 
system
is highly reddened, and believed to consist of a
B1Ia supergiant (Reig et al. 1996) and a neutron star companion.
Since its initial discovery as a binary X-ray source, it has
been observed  by nearly every X-ray experiment that has been launched.
Koenigsberger et al. (1983), carried out a study of the data obtained by the 
OSO-8, 
HEAO 1 and EINSTEIN  observatories, and discovered the presence of X-ray 
flaring
activity on timescales of $\sim$3 hours, in addition to smaller amplitude 
oscillations in the X-ray flux with a period of 894 seconds.
The shorter period was initially identified with the period of the X-ray 
pulsar, but 
Yamauchi et al. (1990) were not able to confirm this  period in data from
the Ginga experiment, finding instead evidence for a posible 850 second 
period.  
On the other hand, the flaring activity has now been shown to be periodic with
P=2.78 hours (Finley et al. 1992; Hall et al. 2000), after combining the data 
from the earlier observations and from ROSAT, EXOSAT, and RXTE.  Finley et al.
(1992) suggested that the 2.78 hour period could actually be 
due to the pulsar's rotation.  This, however, would make it an extremely slow
pulsar (the slowest X-ray pulsar known to date is $\phi$ Per, which has a
period of $\sim$15 minutes), especially when compared with  many of the
X-ray pulsars whose rotation periods are on the order of miliseconds.
Li \& van den Heuvel (1999) showed that the  neutron star could indeed
have reduced its rotation rate within the lifetime of the B-supergiant
companion, but, to do so, it must have had a very large ($>10^{14}$ Gauss)
magnetic field at the time of its birth. Thus, the new-born pulsar would
have been a {\it magnetar}, objects that are predicted theoretically, but
for which few candidates exists as yet (Gotthelf et al. 2002; Marsden et al. 
1999).

An alternative scenario for explaining the periodic X-ray flaring activity
consists of assuming that the optical counterpart undergoes periodic
oscillations and that these oscillations lead to a highly structured
stellar wind.  As the neutron star accretes from shells of stellar wind
with alternating high and low densities, it produces a varying
X-ray flux on the same timescale as the stellar oscillations.  This
suggestion, first presented by Finley et al. (1992), sparked interest
in searching for the 2.78 hour periodicity at optical wavelengths. Although
Taylor et al. (1995) detected 0.009 mag peak-to-peak periodic photometric
variations in the V-band continuum, Finley et al. (1994) and Bell et al. 
(1993)
failed to detect any significant  variations.  This alternative scenario
appears to be less likely than that of the slowly rotating neutron star.
In addition to the difficulty in conclusively establishing  the presence
of periodic variability in the B1-star, there is the problem of being able to
show that the stellar oscillations can lead  not only to a variable
stellar wind, but to a stellar wind that has a {\it periodic} density
structure.

LS {\rm I}+65~010 presents broad and variable H$\alpha$ emission (Margon 1980;
Liu \& Hang 1999; Reig et al. 1996).  Crampton et al. (1985) classified the
optical component as B0.5, based on spectra obtained in 1978 and 1982-84.  
They
noted, however, the difficulty in determining the luminosity class from the
spectrum, since the upper Balmer lines indicate a luminosity class Ia,
but the O, N, and Si lines suggest a class III classification,
while, assuming it is located in the Perseus spiral arm  ($\sim$2.5 kpc),
its luminosity would correspond to class II. Aab et al. (1983) proposed that
the primary in 2S\,0114+650 may be hydrogen-deficient, thus explaining the
weaker strengths of the Balmer lines.  However, Reig et al. (1996)
determined a B1 Ia spectral classification, based on optical spectra and
infrared and optical photometry, a
classification that is supported by Liu \& Hang (1999).  Reig et al. (1996)
also suggest a larger distance to the system of $\sim7\pm3.6$ kpc.
In Table 1 we summarize the  parameters that have been derived for this
system.

The orbital period was determined by Crampton et al. (1985), who
found a very small semi-amplitude (K=17 km s$^{-1}$) for the B-star's
radial velocity variations and they were unable to distinguish between
a circular (P=11.591 days) and an eccentric (e=0.16, P=11.588 days)
solution for the orbital motion.  Recently, Corbet et al. (1999) found
evidence for a slightly longer period, 11.63 days (T$_0$=2450091.36),
than that derived by Crampton et al., based on phase-dependent variations
in the RXTE observations that are interpreted as eclipses of the X-ray
source.  This period, however, is not consistent with the optical RV curve
of Crampton et al., leading Corbet et al. to conclude that the orbital periods
derived from the optical and X-ray measurements are apparently inconsistent,
with no explanation yet available for this discrepancy.

In this paper we present the analysis of optical spectra of LS {\rm I} +65 010
obtained in 1993, 1995 and 2001 with the objective
of studying the B1-star's  spectral variability.  The goals of this
paper are to establish limits on the amplitudes and nature of
line profile variability, and to search for periodicity, on the 2.78 hour 
X-ray
flaring period, which would connect the X-ray behavior with possible 
oscillations
in the B1-star.  The absence of these oscillations would support the
scenario that the neutron star in 2S\,0114+650 was born as a magnetar.

\section{Observations and Data Reduction}

Seven sets of observations of LS {\rm I}+65 010, summarized in Table 2, were 
obtained
between 1993 and 2001. Three sets (1993 October 9,
10 and 13 (UT)) were obtained using the REOSC echelle with  the 2.1 m 
telescope
of the Observatorio Astron\'omico Nacional de San Pedro Martir (OAN/SPM) and 
with
the camera system described by Diego \& Echevarria (1994). A $1024\times1024$ 
Photometrics
CCD detector was used, having pixel size of 19 $\mu$m.   With the 300 grooves 
mm$^{-1}$ 
echellete, the system gives a resolution  of of 7.8 \AA\ mm$^{-1}$ (0.15 \AA\ 
pixel$^{-1}$)
at H $\beta$ and 10.7 \AA\ mm$^{-1}$ (0.20 \AA\ pixel$^{-1}$) at H$\alpha$.  
The resolution is R=17,000, which corresponds  to 17 km s$^{-1}$ per 2
pixels.  The slit was set at a width of 150 $\mu$m, which corresponds to 
2\arcsec.  Each night, a sequence of bias frames was obtained at
the beginning and at the end of the night, as well as sequences
of lamp flat fields.   The target star was observed with 15 or 30
minute exposure times, in a sequence of 3 or more exposures, 
preceded and followed by an exposure of a Th-Ar comparison lamp.
During this observing run, the relative humidity was quite high, 
particularly on the night of October 13, which led to a problem of
condensation on the window of the CCD detector system. The spectral 
region most affected by this problem is the echelle order containing 
H$\alpha$.

Two sets of observations were  obtained on 1995 November 28 and 29 with the 
University
of Hawaii 2.2 m telescope using the f/34 coud\'e spectrograph with
a 900 groove mm$^{-1}$ grating blazed at 4400 \AA , yielding a resolution
of 0.10 \AA\ pixel$^{-1}$. The slit was 1\arcsec\ (350 $\mu$m) wide,
projecting to 2 pixels on an Orbit $2048\times2048$ CCD.   Both
internal-lamp and dome flats were obtained each night as well as a series
of bias frames.  Each exposure of LS {\rm I}+65 010 was of 20 minutes 
duration.
Spectrophotometric standards were observed throughout the night, 
and Th-Ar comparison arc lamps were obtained each night for wavelength
calibration.

Two additional sets were acquired in 2001, one on January 22 and one on 
October 13,
also at the 2.1m of the OAN/SPM.  The obervations of January 22  were made 
using
the CCD Thompson 2048$\times$2048 detector.  The observations of October 13 
were made with
the CCD SITe3.
In Table 2 we list the mean values of the heliocentric Julian date, the 
orbital
phases according to Crampton et al. (1985), the type of observation, the 
number of
individual spectra for each set, the S/N near the He I 5875 \AA~ line, and the
useful wavelength range covered.  In this table we have listed  the
orbital phases according to both the circular and the eccentric orbit 
solutions of
Crampton et al (1985).  However, in the remainder of this paper we will adopt 
the circular solution since it is more consistent with our observations
(see Section 3.3).  It is very important to note that according to the
Crampton et al. circular orbit ephemeris, phase $\phi=0$ corresponds to the
phase at which the B-star is {\it receeding} fastest from the observer.  Thus,
the collapsed object is closest to the observer (i.e., ``in front" of the 
B-star)
at phase $\sim0.25$ and ``behind" the B-star at phase $\sim0.75$.  From
Table 2 we can see that there is  one set of observations (2001 Janunary) 
during
which the collapsed companion was nearly ``behind"  the B-star. According to 
the 
eccentric orbit solution, $\phi$=0 occurs at periastron passage.

Most of the data reduction was performed with IRAF, using the standard
reduction procedures. The extraction and correction for the background of the 
echelle orders of the 2001 October data, however, was made using IDL. The 
images
were corrected with the mean of
the bias frames, and the 1993 and 1995 data sets were also corrected with the
averaged and normalized flat field frames. The 2001 data were not corrected 
for 
flat fields because excessive noise would have been introduced by the flat 
fields.
The orders of the echelle data  were extracted and corrected for the
background.

The wavelength calibration was performed using the comparison lamps (He-Ar) 
that
were obtained nearest in time to the observation. The rms of the polynomial 
fits
to the echelle dispersion functions were in all cases $\leq$ 0.05 \AA. A shift 
in the
zero point of the wavelength scale was applied to each set of spectra to place 
the
principal component of the Na I 5889.95 \AA~ line at its laboratory 
wavelength.  The
velocity shifts needed to accomplish this are listed in column 9 of Table 2. 
Also
listed in this table, in the last column, is the value of $\Delta$V$_{helio}$, 
the
correction for the motion of the Earth with respect to the Sun, calculated 
with
the {\it RVCORR} routine in IRAF. The difference between the values in these
two columns (i.e., col.10$-$col.9), gives the correction that needs to be 
applied
due to the motion of the Earth, to get the heliocentric velocity.
The average value of this correction for the 7 data sets is $-$15 km s$^{-1}$. 
The
conversion to velocities with respect to the LSR require an additional shift 
by
$+$ 7.2 km s$^{-1}$. The velocities listed in the tables are all measured with 
respect
to the laboratory rest wavelengths as are references to velocities in the 
text, except
when specifically noted otherwise.

The internal precision of the wavelength calibration was checked by comparing 
from
spectrum to spectrum the following: 1) the wavelengths of the comparison lamp 
lines;
2) the wavelengths of the [OI] sky lines at 5577.335, 6300.304 and 6363.776 
\AA~ when
observable; and 3) the wavelength of the diffuse interstellar bands (DIBs, see 
Herbig 1995),
particularly the one at $\lambda 5849.65$.  We find that the accuracy of the
wavelength calibration within each data set, for the same line measured on the 
same
echelle order, is 3$-$6 km s$^{-1}$. The accuracy is
largest in the red orders, and in the central portions of each order; it is
lower in the blue orders and on the edges of each order because these portions 
of the
echelle spectrogram have a smaller signal, and the extraction of the spectral 
orders
is less accurate.  In these regions of the spectra, the  uncertainties are 
close to
20 km s$^{-1}$ ($\sim$one resolution element). Because of the large change in
the response of the instrument between the center of the order and the edges, 
it was also
found that the line profiles of photospheric lines that lie near the edges of 
the orders
can change, from one order to the other overlapping order, in the same 
spectrum. Hence,
we concentrate our attention only on lines that lie close to the center of the
order. From the Na I line measurements we find that the internal
precision of the RV measurements near He I 5875 \AA\ is $\pm3$ km s$^{-1}$ for 
the echelle
data and $\pm1$ km s$^{-1}$ for the coud\'e 1995 data.

One of the most serious problems with the echelle data is the rectification of
the orders.  This problem was more severe in the 2001 data than in the 1993 
data.
The instrumental response of each order of the 1993 data was removed
by  fitting a Legendre polynomial (usually of third order) to
each order, for each  spectrum, and dividing the data by this
polynomial.  In  the majority of the spectral orders, this produced a
flat  normalized spectrum.  This is not the case for the order 
containing the line of H$\alpha$ where, due to the humidity problems,
a satisfactory fit to the entire order was not achieved, so that the 
fit was made to the central portions of the order only. The data of 2001 were
rectified using cubic spline functions, in some cases of orders as
high as 15, and even then, the edges of the echelle order could not be fit
in a satisfactory manner.  We estimate that the uncertainty in the equivalent
width measurements in the red, due to the rectification of the orders, is 
$\sim$5\%
near the center of the orders, and 10\% near the edges.  These estimates
were obtained by repeatedly measuring W$_\lambda$'s of the same lines in the 
same echelle
order which, however was rectified  with different functions (cubic
splines and Chebychev) of orders ranging from 7 to 15, and averaging the 
resulting
equivalent width measuremets of lines within the spectral order.

The signal-to-noise ratio (S/N) for the individual 1993  spectra ranges 
from $\sim5$ in the blue to $\sim30$ in the red, for the 1993 October 9 and
10 data, while for the October 13 data, the S/N is somewhat lower.  For the  
1995 data, S/N $\sim35$. The data of 2001 January have S/N $\sim15$ at 3850 
\AA,
increasing with wavelength up to values of 120 in the red on some of the 
individual spectra. The data of 2001 October have S/N$\sim10$ at 4400 \AA\
and $\sim$80 in the red.  Data in the blue were smoothed using Gaussian
($\sigma$=2) or boxcar smoothing functions before the measurements were made.

The line profiles were measured using the deblending routine in IRAF,
fitting one, two, or three Gaussians to the data. In general, one Gaussian
produces a very poor fit to the strong stellar line profiles because they are
not symmetrical.  Lines such as He I 5875 \AA~were fit well with
two Gaussian functions which we will denote as the ``blue" and the ``red" 
Gaussian,
($Vel^B$ and $Vel^R$, respectively) according to their location on the line 
profile.
In general, the ``red" Gaussian corresponds to the line core (i.e., the 
strongest portion
of the line), while the ``blue" Gaussian refers to the line wing on the 
short-wavelength
side of the profile. We also measured the radial velocities using a single 
Gaussian fit
($Vel^{One}$), in order to be able to compare our velocities with those 
obtained
by other authors.

Tables 3-5 list, in column 1, the identifying number of the spectrum, in 
column 2 its
S/N value near He I 5875 \AA, in column 3, the HJD (-2449000 for 1993; 
-2450000 for
1995 and -2451900 for 2001),  and the orbital phase in column 4. In column 5,
we list a phase that was computed using the 2.78 hour X-ray flaring period and 
the
first value of the HJD in the corresponding table as the initial epoch; in 
Columns 6-7,
we list the velocities of the two-Gaussian fits to He I 5875 \AA, and, in 
columns 9-10, the
corresponding equivalent widths (in \AA); in column 8, we list the velocity 
obtained by
measuring the centroid of the line with a single Gaussian fit.  Finally, in 
column 11,
we list the measured velocity of the Na I 5889.95 \AA~ principal ISM 
component.  Absence
of data in these tables is in most cases due to the presence of a cosmic ray 
hit on
the feature. In Table 6 we summarize the results of Tables 3-5 and list the 
average and
standard deviations of the measured parameters.

\section{Results}

\subsection{Spectrum and Interstellar Medium Components}

Figure 1 illustrates the blue portion of the 2001 Jan. spectrum of LS {\rm 
I}+65~010.
As in the spectra obtained by Crampton et al. (1985) and Reig et al. (1996),
there is no evidence for \ion{He}{2} 4686 \AA\ line emission. The only 
emission
line present is H$\alpha$. The other lines of the H-Balmer series are all in 
absorption, as are lines from He I, C III, Si III, O II among other ions.  
Reig et al. (1996) used the ratio of SiIII $\lambda$4552/HeI $\lambda$4387 to
support their supergiant luminosity classification. For a Ia supergiant, this
ratio $\sim$1, decreasing systematically for less luminous stars.  From our 
data of 2001 we obtain SiIII $\lambda$4552/HeI $\lambda$4387$\sim$0.90 (Jan.) 
and $\sim$0.80 (Oct.).  It is interesting to note that Crampton et al.'s 
average 
values for these lines yields SiIII $\lambda$4552/HeI $\lambda$4387$\sim$0.70.
Hence, we find that this (classical) luminosity indicator is somewhat smaller 
than
unity, but may be variable in LS {\rm I}+65 010. 

A new feature that is observable in the 1995 coud\'e data set is
the presence of 3 resolved ISM components in the \ion{Na}{1} D lines, as 
illustrated in 
Figure 2, where both atomic transitions are plotted on a velocity scale.  The 
two high 
velocity components are labeled {\it a} and {\it b}.  The third (principal) 
component 
is too broad to be attributed to a single ISM velocity, so we have assumed 
that it 
consists of at least two unresolved components, labeled  ({\it c} and {\it 
d}).
The four components were de-blended simultaneusly using the IRAF Gaussian
fitting routines, and the radial velocities and equivalent widths of
these features are listed in Table 7. The resulting velocities of these 
components,
are v$_{helio}$ $-$70, $-$48, $-$24, and $-$8 $\pm$3 km s$^{-1}$.  Referred to 
the 
Local Standard of Rest these velocities are 7 km s$^{-1}$ more positive.
The echelle data do not have sufficient spectral resolution to allow the three
individual components a,b, and c+d to be separated as clearly as in the 
Coud\'e
data, and thus, the two high-velocity components are blended together, 
appearing as an
extension towards shorter wavelengths of the principal (c+d) component.  A 
similar
profile is also present in the \ion{Ca}{2} 3934 \AA line.

If we assume that the \ion{Na}{1} D lines are optically thin, we can estimate
the column densities for each of the four velocity components, using
(Spitzer 1978):

\begin{equation}
N(cm^{-2}) = 1.13 \times 10^{17} \frac{W_\lambda (m\AA)}{f \lambda^2 (\AA)}  ,
\end{equation}

\noindent where {\it f} is the oscillator strength.  This assumption is valid
if each observed absorption consists of a blend of un-damped individual lines.
The results are listed in Table 7.  Note that the resulting column densities
for the {\it c,d} components of \ion{Na}{1} $\lambda$5895 are nearly twice
the values derived from the \ion{Na}{1} $\lambda$5889 line, showing that
these components cannot be treated as a blend of un-damped lines, as 
expected from the fact that the absorptions reach zero intensity. Hence, the
derived column densities for these components may be regarded only as lower
limits.  The approximation is more appropriate for each of the two high 
velocity 
components ({\it a} and {\it b}), for which we adopt the value 
N$_{col}=1.4\pm0.3\times10^{12}$ cm$^{-2}$.

A sample of radial velocities of the interstellar lines in the northern Milky 
Way, as a
function of Galactic longitude, is given by M\"unch (1957).  At longitude
{\it l} between 120$^{\circ}$ and 130$^{\circ}$ (LS {\rm I}$+$65 010 lies at
{\it (l,b)}$_{1950}$=(125$^{\circ}$.7,+2$^{\circ}$.6)), he reports a principal 
component at v$_{helio}$ $\sim$ $-$4 km s$^{-1}$, and a secondary components 
at
v$_{helio}$ $\sim$ $-$20 km s$^{-1}$.  Within the uncertainties, these 
velocities  
agree with the velocities we find for components {\it c} and {\it d}. 
The {\it d} component arises primarily in gas that is located in the Orion 
arm,
within 0.5 kpc from the Solar neighborhood, while the {\it c} component arises
in the Perseus spiral arm, which is at $\sim$2.5 kpc. The velocities are 
consistent
with what is expected from the Galactic rotation curve. Our measured 
equivalent
widths, however, are significantly ($\sim$30\%) smaller than those measured by 
M\"unch for the Orion arm regions, while our values for  the Perseus arm 
equivalent widths lie within the range given by M\"unch.

M\"unch does not report the presence of higher velocity components at 
longitudes
near that of LS {\rm I}+65 010. In addition, due to the shape of the Galactic 
rotation
curve (Clemens, 1985), such large negative velocities are excluded, even for 
regions 
that are at greater distances from the Sun.  Hence, we conclude that 
components {\it a} 
and {\it b} are most likely associated with interstellar material in the 
vicinity of 
LS {\rm I}+65 010, and which is expanding away from this system.  We do not 
detect
any variability in the ISM components between our data sets.

\subsection{W$_\lambda$-RV correlation and the systemic velocity}

The equivalent width and the centroid of the unblended photospheric absorption 
lines
with good S/N  were measured on each spectrum of each epoch. The best data 
correspond
to the 2001 January data set, and are listed in Table 8. Column 1 gives the 
laboratory
wavelength of the absorption; column 2, the average velocity (in km s$^{-1}$, 
with respect 
to its laboratory wavelength); column 3, the Gaussian FWHM (in km s$^{-1}$); 
column 4,
the equivalent width (in \AA); column 5 the value of log(gf), where g is the
statistical weight and f the oscillator strength of the transition; and column 
6, 
comments.  All values of log(gf) were taken from the data base available
electronically at the National Institute of Standards and Technology 
({\it http://physics.nist.gov}). For the 2001 October data it was possible to 
measure reliably only about half of the lines (listed in Table 9). 

The data in Tables 8 and 9 are plotted in Figure 3 and suggest the presence of 
a 
correlation between equivalent widths and radial velocity.  In this figure we
plot the data obtained on 2001 January (dark symbols) and October (light 
symbols).  
The H and He lines (triangles and crosses, respectively) observed in 2001 
January
were used to test the reality of the trend. A Spearman's rank correlation 
coefficient
between RV and the EW is r=0.79 with a two-sided significance of 0.001 of its 
deviation 
from zero (small value means a significant correlation). 

Also, the systematic velocity shift of 15  km s$^{-1}$ is evident between 
the data of January (dark symbols) and October (light symbols), which 
coincides with 
the reported semi-amplitude of the radial velocity curve (Crampton et al. 
1985; see below).  
The RVs of the weaker lines are  clustered near $-$20 km s$^{-1}$, and the 
strongest 
lines have RVs $\sim$$-$50 km s$^{-1}$. 

Aab \& Bychkova (1983) and  Crampton et al. (1985) noted the presence in LS 
{\rm I}+65 010 
of a correlation between radial velocity and the excitation of the  atomic 
transition responsible for the line, and this is, generally speaking, the same 
effect 
we are observing in our data. Hutchings (1976, and reference therein) was the 
first to study this correlation in a large sample of stars, and more recent 
analyses have been made by  Massa et al. (1992) and Kudritzky (1992). The 
explanation 
for this effect is that the atmospheres of hot stars are not in hydrostatic 
equilibrium, except perhaps, for the deepest-lying layers. Thus, there is a 
velocity 
gradient in the line-forming regions. Lines that have large opacities, and 
are therefore  observed to arise in exterior atmospheric layers, are shifted 
to shorter
wavelengths than lines that can be observed arising from deeper layers (i.e., 
the optically
thinner lines). In Figure 4 we plot the same H and \ion{He}{1} data shown in 
Figure 3 for
the 2001 January data but use the value of log(gf) in the ordinate, instead of
equivalent width. A very clear correlation is observed, confirming the 
dependence of
the measured velocity on the transition probability of the line.  Given these 
effects, 
it is clear that the  the best approximation to the actual radial velocity of 
the 
B-star is the velocity given by the lines with the smaller log (gf). For the 
2001 
January data, this value is $-$20$\pm$5 km s$^{-1}$, with respect to the 
laboratory reference frame, so v$_{helio}$=$-$35 km s$^{-1}$.

If the photosphere is stable, and its expansion velocity gradient remains 
constant 
throughtout the orbital cycle, then the orbital motion may be well described 
by 
the strong absorption lines. If this is not the case, it is necessary to 
determine 
the orbital parameters exclusively from the weaker lines. 

Adopting the Crampton et al. (1985) orbital solution, our 2001 January data 
are 
very close to a conjunction, and thus the measured radial velocity should  be 
very close to the velocity of the system; i.e., the velocity at $\phi$=0.71 
is $-$4 km s$^{-1}$ with respect to the systemic velocity.  Therefore, the 
velocity of the 
LS {\rm I}$+$65 010 system is v$_{helio}$=$-$31$\pm$5 km s$^{-1}$.  This 
velocity 
coincides with the radial velocity of the  cluster NGC 281 (at {\it 
l}=126$^\circ$) 
and which lies at a distance of 2.9 kpc (Hron 1987).  Hence, assuming  that 
the 
systemic radial velocity is due entirely to the effect of galactic rotation,  
we conclude that  LS {\rm I}+65 010 is likely to be located in the Perseus arm
within $\sim$ 3 kpc from the Sun.  However, it is possible for the system to 
have
a peculiar velocity, given the fact that the neutron star is the result of a
supernova explosion.   Hence, the observed radial velocity may have a 
contribution 
from the peculiar velocity obtained during the SN event.

\subsection{Radial Velocity Variations}

In this section we  concentrate the analysis on He I 5875 \AA~for 
the following reasons: 1) the 7 data sets contain this line, and thus it is 
the only 
line for which a relatively broad orbital phase coverage is available; 2) its  
proximity 
to the ISM Na I D lines and the DIB at 5849 \AA~allows an accurate check on 
the wavelength 
scale; 3) it lies near the center of the echelle order, and this order has a 
good S/N.

In Fig.~5 we reproduce the Crampton et al. (1985) RV data (open triangles), 
plotted 
as a function of the orbital phase (P=11.591; circular orbit solution), upon 
which 
we superpose the mean RVs of \ion{He}{1} $\lambda$5875 (filled squares) 
measured 
with a single Gaussian fit ($Vel^{One}$), taken from Table 6.  
Our data, though scant,  coincide very well with the Crampton et al. RV data.
Note, however, that the velocities of Table 6 are measured with respect to
the laboratory wavelength. Although not specified in the Crampton et al. 
(1985)
paper, their data are most likely heliocentric velocities.  This means that
although the shape of the RV curve determined by our scant data points 
coincides
well with the Crampton et al. RV curve, if heliocentric velocities were 
plotted,
our RV data would be displaced by $\sim$$-$15 km s$^{-1}$.  This is not 
surprising,
since the Crampton et al. RV curve is based on the average velocity of a wide
range of absorption lines (including lines with smaller RVs), while our data 
points refer only to He I 5875 \AA\, which is one of the most blue-shifted 
lines in
the optical spectral range.  Figure 5 also includes the data published by Reig 
et
al (1996) for the \ion{He}{1} 6678 line.  These data follow the Crampton et 
al. 
RV curve  except at phases $\phi$=0.4 - 0.7, when  a large excursion to much 
more 
negative velocities is observed.
 
We also plotted our data using the eccentric orbit ephemeris (P=11.588 days) 
given 
by Crampton et al., and we find that our data are shifted  by $\sim$0.1 
in phase, with respect to the mean RV curve of Crampton et al., plotted with 
this same period.  Hence, the 11.588 day period may now be discarded.
A more complete phase phase coverage is needed in order to further 
improve on the accuracy of the 11.591-day period.

We indicate in Fig.~5 the approximate phase intervals during which the RXTE 
X-ray 
counts (from Corbet et al. 1999) are maximum (0.25-0.75) and minimum 
(0.75-0.25). 
Note that these phase intervals are centered on elongations (i.e., maximum 
approaching and receding velocities of the stars), rather than on  
conjunctions, so 
that the X-ray minimum cannot be attributed to an eclipse of the X-ray source.  
As
pointed out by Corbet et al. (1999), it is difficult to understand why the RV 
curve 
and the X-ray variations on orbital timescales present this inconsistency.

An additional problem arises when we compare the RV values of the \ion{He}{1} 
$\lambda$5875 \AA~line core (Vel$^R$) with the values of Vel$^{One}$ as a 
function
of orbital phase (Fig.~6). Vel$^{One}$ is the radial velocity that is obtained 
from 
the centroid of the whole line profile, while Vel$^R$ is the radial velocity
of the ``red" portion of the line profile (which is actually the strong line 
core),
when it is fit with two Gaussians. If the line profile were to remain constant 
throughout the orbital cycle, both of these velocities whould describe the 
same 
radial velocity curve, except for a constant velocity shift
between them.  However, this is does not occur in LS {\rm I}$+$65 010:  
Vel$^R$ 
and Vel$^{One}$ have similar values near $\phi=0.$ (X-ray minimum), but very 
different 
values near $\phi=0.5$ (X-ray maximum).  This discrepancy is due to the 
presence of a 
much stronger blue wing in the absorption feature at $\phi=0.36$ (see Section 
3.4), which 
leads to a displacement of the line centroid (Vel$^{One}$)  to more negative 
velocities.
The variation of the  \ion{He}{1} 5875 \AA\ blue wing equivalent width 
is displayed in Figure 7, where an increase around  0.36-0.45 is observed. 
Because of the limited number of orbital phases covered by our data, and 
because
we do not have repeat observations at the  same phases, we are unable to 
conclude that the line-profile variations are phase-locked, and therefore 
actually
distort the RV curve.  However, Barsiv et al. (2001) observe a distortion of 
the 
RV curve obtained from the H$\gamma$ line in the optical counterpart of the 
Vela X-1 
that can be traced to line profile variability.  That is, they also detect the 
appearance 
of an extended blue wing on the photospheric absorption.  It is possible also 
that
the RV measurements  of \ion{He}{1} 6678 \AA~ obtained by Reig et al. (1996),
and plotted in Figure 5, may be affected by the appearance of a blue wing.  

\subsection{Line profile variability}

We illustrate the type of line profile variability that occurs  in He I 5875 
\AA\
from  night to night in Figures 8-9.  In Figure 8  we compare the line profile
at phase 0.71, with the profiles at phases  0.89 and 0.36.
The line profile observed at orbital phase $\phi=0.71$ is the strongest and 
most 
symmetrical, among our data sets.  The line 
profile observed at $\phi=0.89$ displays weaker absorption, and presents 
indications of 
the presence of emission superposed at least on its red  wing. At this phase, 
the neutron
star is near the elongation at which it has the maximum approaching velocity, 
and any
emission arising in its vicinity would be expected to be blue-shifted.  It is 
possible
that the reduced strength of the photospheric absorption is a result of the 
line being
filled-in by such emission.  The profile at $\phi=0.36$  also presents 
indications of 
emission on the red wing, but, in addition, the blue wing has excess 
absorption.  

The H$\alpha$ line is available in only 5 of the 7 data sets, but due to 
severe
fringing problems in the 2001 Jan data within this spectral region, we are not 
able
to reliably use this H$\alpha$  line profile. A montage of the line
profiles is presented in Figure 10.  At orbital phase 0.07 H$\alpha$ presents
an emission-line profile upon which a prominent absorption feature is
superimposed, shortward of the line center. At phase 0.15, the absorption is 
weaker, and
by phase 0.41 it has been replaced by emission. A similar change in this line 
on the 
same timescale is reported to have occurred in 1984 by Minarini et al. (1994). 
The  average (for each night) equivalent width of the emission component does 
not change appreciably, having a value of $W_\lambda=1.2\pm0.1$ \AA, very 
similar to 
the value obtained by Reig et al.(1996). Variations mainly on the blue wing 
of the line profile of H$\alpha$ have also been reported by Liu \& Hang 
(1999), who favor 
contamination by emission, due to the presence of an ionized bubble 
surrounding the neutron 
star.  This ionized region would be expected to partake in the orbital motion 
of the 
neutron star.  Hence, the extra emission would tend to "fill in" the blue 
wing of the photospheric absorption lines only when it is approaching the 
observer.   According to the radial velocity curve of Crampton et al., this is 
at 
$\phi$$\sim$0.80-1.20.  In Figure 10 we do observe a weak emission centered at
$-$270 km s$^{-1}$ at $\phi$=0.07, while at  $\phi$=0.41, there is some 
indication
of excess emission at $+$200 km s$^{-1}$.  However, the line profiles are too 
noisy 
for these excess emissions to be quantified with precision. At phase 0.41
there is no trace of the photospheric absorption, implying that it is 
completely
filled in by the emission.   Thus, the H$\alpha$ line profiles do  provide 
support 
for the hypothesis of an H\,II region surrounding the neutron star.  However, 
if this
were the only source for the emission, a much stronger red-shifted emission 
whould
be present at phase 0.41 than is observed.  Hence, we conclude that there
is a persistent component to the H$\alpha$ emission centered at $\sim$$+$80 km 
s$^{-1}$
that does not follow the orbital motion of the neutron star, and that most 
likely arises 
in the wind of the B1-star.

\subsection{Variability on the 2.78 hour X-ray flaring period}

We searched for variability with $P=2.78$ hours in all the radial velocity 
and equivalent width measurements of \ion{He}{1}.  We were unable to 
detect any convincing modulation in any of the data sets, although variations
of up to 40 km s$^{-1}$ in the ``blue" (Vel$^B$) 2-Gaussian fit to the data
were detected within a single night. 
In Figures 11-12 we plot the velocity of
the ``red" Gaussian (that corresponds to the core of the line)  as a function 
of
phase calculated for the 2.78 hour period, for the  nights 1995 Nov. 28, 29,
(Fig.11) and 2001 Jan 22, Oct 13 (Fig. 12).  The uncertainties in these 
measurements 
are $\sim$10 km s$^{-1}$.  The 2001 data (orbital phases 0.71 and 0.89) are 
practically
constant throughout the night. There is a moderate ($\sim$15 km s$^{-1}$) 
amount of 
variability in the 1995 Nov. data (orbital phases 0.36 and 0.45), but with 
little
evidence indicating that this variability follows the 2.78 hour period. 
The degree of variability is largest for the blue wing of \ion{He}{1} 5875 
\AA\
in the 1995 data sets (Figure 13), although a possible modulation on the 2.78 
hour period that is observed on the Nov. 29 data depends on  3 data points, 
and 
thus can only be considered as marginal evidence in favor of the presence of 
this modulation.  

\section{Conclusions}

We  present the results obtained from optical spectral observations of the 
X-ray 
binary system 2S0114 +650=LS {\rm I}+65 010.
The radial velocity curve obtained from our data are inconsistent with 
the eccentric orbit solution given by Crampton et al. (1986), and thus we 
favor the
P=11.591 days  solution, which assumes a circular orbit.  

The  correlation between the equivalent width of the lines and their radial 
velocity is used to estimate the systemic radial velocity.  Strong lines have 
the largest 
negative velocities because they form further out in the atmosphere, where
expansion velocities are greater than in the inner atmospheric layers. Hence, 
the weaker
lines provide a more accurate estimate of the actual radial velocity of the 
B1-star, which
we find to be V$_{helio}$=$-$31 km s$^{-1}$.  Assuming that the radial 
velocity 
is due to Galactic rotation leads us to conclude that LS {\rm I}+65 010 is at 
$\sim$ 
3 kpc.  This distance is near  the lower limit of the distance of 7$\pm$3.5 
kpc 
derived by Reig et al. (1996). Caution is necessary, however, because the  SN 
event that 
occurred in this system  might have given it a peculiar velocity with respect 
to
its expected velocity due to the Galactic rotation curve. 
The  spectral luminosity indicator SiIII $\lambda$4552/He I $\lambda$4387
$\sim$0.9 in our data, while in the Crampton et al. data it is 0.7, both
values consistent with a high luminosity for the B1-star.  If it is a 
supergiant
($\lambda$4552/He I $\lambda$4387$\sim$1), then a larger distance is expected.

Significant line profile variability is detected from night to night.  The
\ion{He}{1}$\lambda$5875 photospheric line presents an additional, 
blue-shifted 
absorption component at orbital phases 0.36 and 0.45, thus enhancing the 
generally
present asymmetry of this line. The extension of this blue wing is  $-400$ km 
s$^{-1}$, 
far in excess of the broadening that could be attributed to the 100 km 
s$^{-1}$ rotational 
velocity of the B1-star, and indicates that these wings form in portions of 
the photosphere 
that are not only expanding, but where the expansion may depend on the orbital 
phase. 
A similar blue-shifted component to this absorption line appears in the 
spectra of
HD 7751, the optical counterpart of Vela X-1 (Barziv et al. 2001), at orbital 
phases
0.45-0.67, which in this case correspond to phases when the neutron star in 
``in front"
of the B-star primary.  In LS {\rm I}+65 010, phases 0.36-0.45 are closer to 
an elongation,
although the neutron star is still on the near side (with respect to the 
observer) of 
the B1-star. We note that the X-ray maximum occurs at phase 0.5 in LS {\rm 
I}+65 010, so there
may be a connection between the appearance of the extra blue-shifted 
absorption and the
enhanced X-ray emission.  Our data, however, are insufficient to be able to 
conclude
that the observed variability is phase-locked, although this conclusion was 
reached
by Barziv et al. (2001) for the case of Vela X-1.

Significant variability on an hourly timescale was detected only for two of
our data sets (1995; orbital phases 0.36 and 0.45).  However, we are unable to
conclude that this variability is periodic, on the 2.78 X-ray flaring period.  
It is
clear, however, that the strongest variability occurs on the blue portions of 
the
line profiles, consistent with the notion that the  photospheric layers at the 
base
of the wind are not undergoing a steady expansion, at least at these orbital 
phases
where the variability is detected.  Hence, the effect of the neutron star on 
the
companion's atmosphere may  not be negligible.

Two high velocity components (V$_{helio}=-$70 and $-$49 km s$^{-1}$) are 
detected in the Na I D lines that may be associated with  a circumstellar H II 
region 
surrounding LS {\rm I}+65 010.  This region would have to have formed after
the SN explosion, and thus suggests possible shell ejections from the B-star.
However, we cannot at this stage discard intervening high velocity clouds as
the source of these absorptions.

The behavior of the RXTE X-ray light curve obtained by Corbet et al. (1999)
cannot be explained if the Crampton et al. RV curve represents the true 
orbital motion of the B star. We have tested various scenarios, such as the
presence of a disc, an eccentric orbit and a non-spherically symmetric wind,
and all of these scenarios demand that X-ray minimum occur when the
neutron star is behind the B1-star (phase 0.75); or
the presence of two maxima, assuming the orbit of the neutron star intersects
a high density disk surrounding the B1-star. None of these correspond to the
observations. The only contribution we can make towards trying to solve this
dilemma is the finding that X-ray maximum appears to be correlated with the
presence of a more extended blue wing in some of the photospheric absorption
lines. Comparing the He I 5875 line profiles of LS {\rm I}+65 010 with the 
same line
profiles in Vela X-1 (Fig. 17 of Barziv et al. 2001), and assuming that 
similar
processes occur in both binary systems, it is tempting to speculate that the
Crampton et al. (1985) radial velocity curve may be distorted in the phase
range $\sim$0.3-0.5, such that at $\phi$$\sim$0.5 the neutron star is  
actually
``in front" of the B1-star, rather than at an elogation.  Clearly, however,
an understanding is necessary of the processes involved in producing the 
line-profile variability before this speculation can be taken a step further.
In addition, it would be most desirable to obtain radial velocity measurements
throughout the orbital cycle of 2S0114+650 from the upper Balmer lines and
weak metal lines, all of which are much less affected by the non-stationary
nature of the outer expanding photosphere.

\acknowledgements
We are grateful to an anonymous referee for the meticulous revision of the 
submitted versions, resulting in a much improved paper. We thank John Dvorak, 
Salvador Monrroy, and J. Velazco
for their assistance with the spectroscopic observations. GK thanks M.
Peimbert and D. Massa for enlightening discussions. We thank Elfego 
Ruiz and Daniel Pe\~na for technical advice, and the University of Hawaii for
granting observing time. Extensive use was made of the atomic lines data bases
available through the National Institute of Standards and Technology (NIST).
Part of this work was performed under the auspices of 
the U.S. Department of Energy, National Nuclear Security Administration 
by the University of California, Lawrence Livermore National Laboratory 
under contract No. W-7405-Eng-48, and and grants from CONACYT and 
from UNAM/DGAPA.

\clearpage

\begin{table}[!t]\centering
 \setlength{\tabnotewidth}{0.9\textwidth}
  \tablecols{3}
\caption{Parameters of LS {\rm I}+65 010 = 2S\,0114+650}
\begin{tabular}{lcc}
\toprule

Parameter & Value & Notes and references \\

 \midrule

   Spectral type  &    B1\,Ia           &                    5 \\
   T$_{eff}$      &    $24000\pm3000$   &                    5 \\
   R/R$_{\odot}$  &    $37\pm15$        &                    5 \\
   M/M$_{\odot}$  &    $15\pm5$         &                    5 \\
   {\rm M$_V$}    &    $-7.0\pm1.0$     &                    5 \\
   Distance (kpc) &    $7.0\pm3.5$      &                    5 \\
                  &    2.5              &                    2 \\
  {\it vsini} (km/s) & $96\pm20$; 46    &                    5,2; 7\\
   E(B-V)         &    $1.24\pm0.02$    &                    5 \\
   log g          &    $2.5\pm0.2$      &                    5 \\
   BC             &    $-2.3\pm0.3$     &                    5 \\
   {\rm M$_{bol}$}&    $-9.3\pm1.0$     &                    5 \\
   Orbital period (d) & $11.588\pm0.003$ &    e=0.16;        2 \\
                    &   $11.591\pm0.003$ &    e=0;           2 \\
   Initial Epoch T$_o$ & $2444134.9\pm0.7$  &   e=0.16;   2   \\
                      & $2444134.3\pm0.2$ &     e=0;     2  \\
   K (km s$^{-1}$)  & $17\pm1$ &                            2\\
   L$_x$(1.5-10 kev)  & $1.3\times10^{35}$    & d=2.5 kpc; 2 \\
                      & $1.1\times10^{36}$    &              6 \\
   P$_{pulse}$ (secs)    &  893.8 ?          &    1 \\
                    &  850             &    3 \\
   P$_{X-flares}$ (hrs)   &    2.78          &    4 \\
                        &    2.73          &    6 \\
   T$_{o,X-flares}$     & $2446433.632\pm0.06$ &    4 \\
\bottomrule

\tabnotetext{}{ References for Table 1: 1) Koenigsberger et al. 1983;
2) Crampton et al. 1985; 3) Yamauchi et al. 1990; 4) Finley et al. 1992;
5) Reig, et al. 1996; 6)  Hall et al. 2000; 7) Aab et al. 1983.
T$_o$ corresponds to periastron passage in the eccentric orbit and
   maximum positive velocity in circular orbit.}
 \end{tabular}
\end{table}
\clearpage

\begin{table}[!t]\centering
  \setlength{\tabnotewidth}{0.9\textwidth}
  \tablecols{10}
  \setlength{\tabcolsep}{\tabcolsep}

\caption{Summary of the Observations}
\begin{tabular}{lllllrlllr}
    \toprule
Year & mean HJD & $\phi^{circ}$ & $\phi^{ecc}$ &Type &Num. &S/N  & 
$\lambda$-range & zero-point  & $\Delta$V$_{helio}$   \\
 & -2400000& & & & &at $\lambda5875$ &\AA&shift km s$^{-1}$ & km s$^{-1}$ \\

\midrule
1993 & 49269.943 & 0.07 & 0.13 & SPM echelle  &    5      &     26-33 
&4300-6650 &+22& 9.5 \\
1993 & 49270.834 & 0.15 & 0.21 & SPM  echelle &    10     &     16-37 
&4300-6650 &+22& 9.1 \\
1993 & 49273.894 & 0.41 & 0.47 & SPM  echelle &    14     &     17-27 
&4420-6600 &+20& 8.4 \\
1995 & 50049.882 & 0.36 & 0.44 & UH  coud\'e  &    13     &     30-45 
&5815-5956 &+10&-5.9 \\
1995 & 50050.884 & 0.45 & 0.53 & UH  coud\'e  &    10     &     20-47 
&5815-5956 &+ 8 &-6.1\\
2001 & 51931.682 & 0.71 & 0.83 & SPM echelle  &    11     &     
50-120&3700-7500 &- 1 &-18.0\\
2001 & 52095.858 & 0.87 & 0.00 & SPM echelle  &    14     &     40-80 
&4350-6850 &+26 &8.4 \\
    \bottomrule
  \end{tabular}
\end{table}\clearpage

\begin{table}[!t]\centering
  \setlength{\tabnotewidth}{0.9\textwidth}
  \tablecols{11}
  \setlength{\tabcolsep}{\tabcolsep}
\caption{Echelle 1993 Observations of He I 5875}

\begin{tabular}{lcccccrcccc}
    \toprule

ID & S/N&HJD&$\phi^{cir}$ &$\phi^{flare}$&Vel$^B$&Vel$^R$ & 
Vel$^{One}$&W$_\lambda^B$&W$_\lambda^R$&NaI  \\
(1) & (2)&(3) &(4) &(5) &(6) &(7) &(8) &(9)&(10)&(11)\\

\midrule
     3  &   20.  & 49269.860  &  0.065  &  0.000  &-260 & -43 & -46 & 
0.16&1.00 & 0  \\
     7  &   25.  & 49269.918  &  0.070  &  0.501  &-200 & -44 & -52 & 
0.25&0.92 & -3 \\
     9  &   27.  & 49269.947  &  0.072  &  0.751  &-197 & -41 & -46 & 
0.15&0.89 & 0 \\
    11  &   32.  & 49270.000  &  0.077  &  0.209  &-209 & -45 & -50 & 
0.11&0.93 & -1 \\
    13  &   25.  & 49270.026  &  0.079  &  0.434  &-197 & -43 & -47 & 
0.14&0.81 & -2 \\
     6  &   32.  & 49270.712  &  0.138  &  0.358  &-246 & -39 & -40 & 
0.10&0.91 & 1 \\
     8  &   26.  & 49270.740  &  0.141  &  0.599  &-254 & -38 & -40 & 
0.12&0.91 & 0 \\
    10  &   33.  & 49270.765  &  0.143  &  0.815  &---  & --- & --- & --  & --  
& -1 \\
    12  &   32.  & 49270.792  &  0.145  &  0.048  &-234 & -38 & -43 & 
0.20&0.90 & -2 \\
    14  &   30.  & 49270.818  &  0.147  &  0.273  &-233 & -38 & -38 & 
0.13&0.90 & 0 \\
    16  &   20.  & 49270.860  &  0.151  &  0.636  &---  & --- & --- & --  & --  
& -4 \\
    18  &   26.  & 49270.885  &  0.153  &  0.852  &-239 & -39 & -44 & 
0.19&0.92 & -3 \\
    22  &   26.  & 49270.932  &  0.157  &  0.257  &-216 & -37 & -44 & 
0.19&0.80 & -4 \\
    23  &   26.  & 49270.943  &  0.158  &  0.352  &-252 & -41 & -42 & 
0.16&0.89 & -5 \\
    24  &   20.  & 49270.957  &  0.159  &  0.473  &-245 & -41 & -43 & 
0.12&0.91 & -3 \\
     6  &   17.  & 49273.774  &  0.402  &  0.800  &-162 & -28 & -47 & 
0.10&0.96 &  3  \\
    10  &   21.  & 49273.840  &  0.408  &  0.370  &-178 & -29 & -49 & 
0.27&0.74 &  0 \\
    11  &   23.  & 49273.851  &  0.409  &  0.465  &-195 & -38 & -46 & 
0.17&0.77 & -3 \\
    13  &   15.  & 49273.874  &  0.411  &  0.663  &---  & --- & --- & --  & --  
& -2 \\
    14  &   19.  & 49273.886  &  0.412  &  0.767  &-171 & -31 & -53 & 
0.25&0.79 & -3 \\
    15  &   16.  & 49273.897  &  0.413  &  0.862  &---  & --- & --- & --  & --  
& -3 \\
    16  &   14.  & 49273.907  &  0.414  &  0.948  &-276 & -50 & -58 & 
0.20&0.94 & -5 \\
    18  &   22.  & 49273.936  &  0.416  &  0.199  &-215 & -36 & -50 & 
0.24&0.80 & -3 \\
    19  &   19.  & 49273.947  &  0.417  &  0.294  &---  & --- & --- & --  & --  
& -- \\
    20  &   14.  & 49273.958  &  0.418  &  0.389  &---  & --- & --- & --  & --  
& -5 \\
    21  &   13.  & 49273.980  &  0.420  &  0.579  &---  & --- & --- & --  & --  
& -5 \\
    22  &   15.  & 49273.991  &  0.421  &  0.674  &-203 & -36 & -58 & 0.29& 
0.90& -5 \\
    23  &   18.  & 49274.003  &  0.422  &  0.777  &-222 & -51 & -56 & 0.21& 
0.77& -5 \\
    24  &   13.  & 49274.015  &  0.423  &  0.881  &---  & --- & -62:& --  & --  
& -8 \\
    \bottomrule
  \end{tabular}
\end{table}\clearpage

\begin{table}[!t]\centering
  \setlength{\tabnotewidth}{0.9\textwidth}
  \tablecols{11}
  \setlength{\tabcolsep}{\tabcolsep}

\caption{Coud\'{e} 1995 Observations of He I 5875 }
\begin{tabular}{lcccccrcccc}
\toprule
ID & S/N&HJD&$\phi^{cir}$&$\phi^{flare}$&Vel$^B$&Vel$^R$ & 
Vel$^{One}$&W$_\lambda^B$&W$_\lambda^R$&NaI  \\
(1) & (2)&(3) &(4) &(5)&(6) &(7) &(8) &(9)&(10)&(11) \\

\midrule

    21&   31.& 50049.802&  0.353&  0.000& -176.&  -27.&  -73.&  0.60&  0.56&    
1.\\
    22&   30.& 50049.816&  0.354&  0.121& -195.&  -31.&  -65.&  0.41&  0.77&    
1.\\
    23&   33.& 50049.828&  0.355&  0.224& -166.&  -21.&  -76.&  0.61&  0.56&    
0.\\
    24&   33.& 50049.839&  0.356&  0.319& -166.&  -23.&  -71.&  0.58&  0.60&    
0.\\
    25&   35.& 50049.851&  0.357&  0.423& -192.&  -36.&  -80.&  0.55&  0.62&   
-1.\\
    26&   36.& 50049.863&  0.358&  0.526& -165.&  -29.&  -79.&  0.60&  0.54&    
0.\\
    28&   40.& 50049.879&  0.360&  0.664& -173.&  -25.&  -73.&  0.58&  0.62&    
0.\\
    29&   43.& 50049.894&  0.361&  0.794& -181.&  -30.&  -70.&  0.38&  0.64&    
1.\\
    30&   45.& 50049.908&  0.362&  0.915& -186.&  -29.&  -79.&  0.46&  0.66&    
0.\\
    31&   44.& 50049.923&  0.363&  1.044& -195.&  -36.&  -65.&  0.30&  0.69&    
0.\\
    32&   36.& 50049.937&  0.365&  1.165& -204.&  -27.&  -68.&  0.48&  0.74&    
1.\\
    33&   34.& 50049.954&  0.366&  1.311& -218.&  -36.&  -75.&  0.35&  0.69&    
1.\\
    34&   35.& 50049.968&  0.367&  1.432& -190.&  -25.&  -67.&  0.39&  0.68&    
1.\\
    97&   47.& 50050.780&  0.437&  8.438& -222.&  -48.&  -59.&  0.28&  0.77&    
0.\\
    98&   43.& 50050.796&  0.439&  8.576& -221.&  -50.&  -64.&  0.29&  0.77&    
1.\\
    99&   40.& 50050.810&  0.440&  8.697& -223.&  -50.&  -58.&  0.22&  0.79&    
0.\\
   100&   27.& 50050.825&  0.441&  8.827& -220.&  -48.&  -66.&  0.26&  0.72&    
0.\\
   101&   23.& 50050.841&  0.443&  8.965& -231.&  -57.&  -75.&  0.30&  0.85&    
0.\\
   102&   24.& 50050.855&  0.444&  9.085& -240.&  -56.&  -84.&  0.37&  0.74&    
0.\\
   103&   35.& 50050.870&  0.445&  9.215&\nodata& -46.&  -60.&  0.22&  0.73&    
0.\\
   104&   23.& 50050.885&  0.446&  9.344& -233.&  -46.&  -55.&  0.28&  0.77&    
0.\\
   106&   37.& 50050.915&  0.449&  9.603& -222.&  -55.&  -67.&  0.26&  0.75&    
0.\\
    \bottomrule
  \end{tabular}
\end{table}\clearpage

\begin{table}[!t]\centering
  \setlength{\tabnotewidth}{0.9\textwidth}
  \tablecols{11}
  \setlength{\tabcolsep}{\tabcolsep}

\caption{Echelle 2001 Observations of He I 5875 }
\begin{tabular}{lcccccrcccc}
\toprule
ID & S/N&HJD&$\phi^{cir}$&$\phi^{flare}$&Vel$^B$&Vel$^R$ & 
Vel$^{One}$&W$_\lambda^B$&W$_\lambda^R$&NaI  \\
(1) & (2)&(3) &(4) &(5) &(6) &(7) &(8) &(9)&(10)&(11)\\
\midrule
  4020&   80.& 51931.616&  0.705&  0.004& -228.&  -58.&  -63.& 0.12 &1.14  &   
-1.  \\
  4021&  120.& 51931.627&  0.705&  0.099&\nodata& -62.&  -65.& 0.06 &1.23  &   
-1.  \\ 
  4023&   79.& 51931.641&  0.707&  0.220& -287.&  -59.&  -66.& 0.10 &1.17  &   
-1.  \\
  4024&   76.& 51931.652&  0.708&  0.315& -342.&  -58.&  -62.& 0.06 &1.17  &   
-1.  \\ 
  4025&  100.& 51931.664&  0.709&  0.419& -203.&  -62.&  -66.& 0.08 &1.15  &    
0.  \\ 
  4026&   85.& 51931.676&  0.710&  0.522& -274.&  -58.&  -68.& 0.18 &1.24  &    
1.  \\ 
  4028&   87.& 51931.690&  0.711&  0.643& -237.&  -59.&  -63.& 0.08 &1.18  &    
0.  \\ 
  4030&   53.& 51931.713&  0.713&  0.842& -128.&  -66.&  -66.& 0.05 &1.14  &   
-7.  \\ 
  4031&   67.& 51931.725&  0.714&  0.946& -270.&  -63.&  -63.& 0.05 &1.16  &   
-2.  \\ 
  4032&   95.& 51931.736&  0.715&  0.041& -267.&  -60.&  -65.& 0.07 &1.17  &    
2.  \\
  4033&   80.& 51931.748&  0.716&  0.144& -211.&  -60.&  -61.& 0.07 &1.16  &    
3.  \\ 
  1066&   80.& 52095.776&  0.867&  0.621& -214.&  -39.&  -43.& 0.04 &0.68  &    
0.  \\
  1067&   75.& 52095.787&  0.868&  0.716& -133.&  -38.&  -39.& 0.05 &0.71  &    
2.  \\
  1069&   67.& 52095.804&  0.870&  0.863& -227.&  -35.&  -42.& 0.04 &0.67  &    
1.  \\
  1070&   64.& 52095.815&  0.871&  0.958& -214.&  -38.&  -41.& 0.05 &0.68  &    
1.  \\
  1071&   80.& 52095.826&  0.872&  0.053& \nodata& \nodata & \nodata &\nodata 
&\nodata  & -1.\\
  1073&   77.& 52095.844&  0.873&  0.208& -233.&  \nodata &  \nodata &\nodata 
&\nodata &  1.  \\
  1075&   62.& 52095.866&  0.875&  0.398& -167.&  \nodata &  \nodata & 0.03 
&0.63 & -5.\\
  1077&   50.& 52095.881&  0.876&  0.528& -268.&  -41.&  -41.&0.01  & 0.68 &    
3.  \\
  1078&   77.& 52095.892&  0.877&  0.623& -214.&  -36.&  -39.&0.01  & 0.68 &    
2.  \\
  1079&   72.& 52095.903&  0.878&  
0.718&\nodata&\nodata&\nodata&\nodata&\nodata& 5. \\
  1081&   50.& 52095.918&  0.879&  0.847&\nodata &\nodata 
&\nodata&\nodata&\nodata& 5.\\
  1083&   65.& 52095.940&  0.881&  0.037& -188.&  -34.&  -38.& 0.06 &0.70  &    
4.  \\
    \bottomrule
  \end{tabular}
\end{table}\clearpage

\begin{table}[!t]\centering
  \setlength{\tabnotewidth}{0.9\textwidth}
  \tablecols{11}
  \setlength{\tabcolsep}{\tabcolsep}

\caption{Summary of the properties of He I $\lambda$5875  }
\begin{tabular}{lcccrcccr}
\toprule

ID & HJD&$\phi^{cir}$&Vel$^B$ &Vel$^R$ & Vel$^{One}$ &W$_\lambda^B$ 
&W$_\lambda^R$ &NaI  \\
(1) & (2) &(3) &(4) &(5) &(6) &(7) &(8) &(9) \\
\midrule
1993 Oct. 9&49269.943 & 0.07 &  $-213\pm27$ & $-43\pm1$ & $-48\pm3$ 
&$0.16\pm0.05$&$0.91\pm0.07$& $-1\pm1$ \\
1993 Oct. 10&49270.834 & 0.15& $-240\pm12$ & $-39\pm1$ & $-42\pm2$ 
&$0.15\pm0.04$&$0.89\pm0.04$& $-2\pm2$ \\
1993 Oct. 13&49273.894 & 0.41& $-203\pm36$ & $-37\pm9$ & $-53\pm6$ 
&$0.22\pm0.06$&$0.83\pm0.08$&  $-3\pm3$ \\
1995 Nov. 28&50049.882 & 0.36& $-185\pm16$ & $-29\pm5$ & $-72\pm5$ 
&$0.48\pm0.11$&$0.64\pm0.07$& $0.3\pm0.7 $ \\
1995 Nov. 29&50050.884 & 0.45& $-226\pm7 $ & $-51\pm4$ & $-65\pm9$ 
&$0.27\pm0.04$&$0.77\pm0.04$& $ 0\pm0.3$\\
2001 Jan. 22&51931.682 & 0.71& $-230\pm77$ & $-60\pm3$ & $-64\pm2$ 
&$0.08\pm0.03$&$1.17\pm0.03$& $-1\pm3$ \\
2001 Oct. 13&52095.858 & 0.87& $-173\pm63$ & $-37\pm2$ & $-40\pm2$ 
&$0.04\pm0.02$&$0.68\pm0.02$& $ 2\pm3$ \\
    \bottomrule
  \end{tabular}
\end{table}\clearpage

\begin{table}[!t]\centering
   \newcommand{\DS}{\hspace{6\tabcolsep}} 
 \setlength{\tabnotewidth}{0.9\textwidth}
  \tablecols{9}
  \setlength{\tabcolsep}{\tabcolsep}

\caption{Na I ISM components}
\begin{tabular}{lrrrr @{\DS} rrrr}
\toprule

& \multicolumn{4}{c}{Na I $\lambda$5889.950} &  \multicolumn{4}{c}{Na I 
$\lambda$5895.924}\\
\cline{2-5} \cline{6-9}
& \it a & \it b &\it c &\it d & \it a & \it b &\it c &\it d\\

\midrule

f$_{\it ik}$ &  \multicolumn{4}{c}{0.647}  & \multicolumn{4}{c}{0.322} \\
V  (km s$^{-1}$)    &$-$55& $-$34& $-$9&$+$8 &  $-$55 & $-$33 & $-$8 &$+$7\\
V$_{helio}$ (km s$^{-1}$)  &$-$70& $-$49& $-$24&$-$7 &  $-$70 & $-$69 & $-$23 
&$-$8\\
W$_{\lambda}$ (\AA) &0.215 &0.210& 0.330 &0.315 &0.177& 0.172& 0.300& 0.300 \\
N$_{col}$ (10$^{12}$cm$^{-2}$)&1.08 & 1.06 & $>$1.66 & $>$1.59 &1.77& 1.74 
&$>$3.03 &$>$3.03  \\
    \bottomrule
  \end{tabular}
\end{table}\clearpage

\begin{table}[!t]\centering
  \setlength{\tabnotewidth}{0.9\textwidth}
  \tablecols{6}
  \setlength{\tabcolsep}{\tabcolsep}

\caption{Mean radial velocities and equivalent widths in January 2001}
\begin{tabular}{lrrrrl}
\toprule

Line ID & RV &GFWHM& W$_\lambda$ &log(gf) &Comments\\
 & km s$^{-1}$ &km s$^{-1}$&\AA & & \\
 (1)&(2)&(3) & (4) & (5) & (6)\\

 \midrule

 H I       &  &  &  &  & \\
  4101.74          & -34  & 280  & 0.90  & -1.108 &   \\
  4340.47          & -46  & 220  & 1.12  & -0.796 &      \\
  4861.33         & - 55  & 230  & 1.20  & -0.358 &      \\
  6562.82         & -117  & 160  & 0.60  & 0.144 &   abs. superposed on 
emission\\
 HeI       &  &  &  &  &  \\
  3926.53         & -19  & 275   & 0.38  & -1.647 &     seen only in mean \\
  3964.73         & -25  & 130   & 0.30  &  -1.295  &     \\
  4009.27         & -23  & 140   & 0.28  &   -1.473  &     \\
  4026.28         & -42  & 190   & 0.73  & -0.701  &     blend 
4026.19+4026.36\\
  4143.76         & -33  & 170   & 0.45&    -1.196 &         \\
  4387.93         & -37  & 160   & 0.50   & -0.883 &     blend w/ C III 
4388.02\\
  4471.58         & -52  & 170   & 0.93  &  -0.278 &     blend 
4471.48+4471.69\\
  4713.26         & -34  & 130   & 0.35 &   -1.230 &     blend 
4713.15+4713.38\\
  4921.93         & -33  & 150   & 0.79&    -0.435 &     \\
  5015.68         & -33  & 135   & 0.42&    -0.820 &     \\
  5047.74         & -22  & 128   & 0.24  &  -1.601 &     blended w/ N II 
5045\\
  5875.80         & -57  & 180   & 1.25 &   0.408  &     blend:5875.62+ 
5875.97\\
  6678.15         & -46  & 150   & 1.18&    0.329  &     \\
    \bottomrule
  \end{tabular}
\end{table}\clearpage

\begin{table}[!t]\centering
  \setlength{\tabnotewidth}{0.9\textwidth}
  \tablecols{6}
  \setlength{\tabcolsep}{\tabcolsep}

\caption{Cont.}
\begin{tabular}{lrrrrl}
\toprule

Line ID & RV &GFWHM& W$_\lambda$ &log(gf) &Comments\\
 & km s$^{-1}$ &km s$^{-1}$&\AA & & \\
 (1)&(2)&(3) & (4) & (5) & (6)\\

 \midrule

 C III              &  &  &  &  \\
  4388.02         & -46  & 160   & 0.51   & -0.506 & strong line  \\
  5695.92         & -7  & 140   & 0.18    & 0.0170 &  \\
 N II        &  &  &  &  \\
  3995.00         & -42  & 110   & 0.30   & 0.208  & \\
  5045.099        & -32  & 170   & 0.18   & -0.407 &     blended w/ He I\\
  5676.02         & -17  & 170   & 0.18   &  -0.367 & \\
  5679.56         & -37  & 140   & 0.33   &  0.250  & \\
 O II            &  &  &  &  \\
  4075.86        & -40  & 130   & 0.20  & 0.693 &\\
  4319.63        & -31  & 120   & 0.21 &  -0.380 & blended w/ line above \\
  4345.56        & -30  & 120   & 0.21 &  -0.346  &  \\
  4366.90        & -43  & 140   & 0.22&   -0.348  &  \\
  4414.91        & -29  & 160   & 0.25  & 0.172   & blended w/ line below \\
  4416.98        & -32  &  80   & 0.12 &  -0.077  &  blended w/ line above\\  
  4590.97        & -35  & 112   & 0.19&   0.350  &     \\
  4661.64        & -28  & 130   & 0.24 &  0.278  &    \\
  4699.21        & -25  & 140   & 0.08& 0.270     &  \\
 Si III         &  &  &  &  \\
  4552.62         & -36 & 130  & 0.45  &  0.292   &         nice symmetrical line\\
  4567.83         & -40 & 120  & 0.38  &  0.070    &                     \\
  4574.76         & -35 & 130  & 0.26  &  -0.406  &                        \\
  4819.74         & -35 & 180  & 0.24  &  na     &                       \\
  5739.73         & -20 & 170  & 0.35  & -0.157 &                         \\
Si IV          &  &  &  &  \\
  4116.10    & -25 & 110  & 0.16  &     -0.106 &                    \\
Mg II         &  &  &  &  \\
  4481.23    & -51 & 160  & 0.23  &   0.730   & blend 4481.13+4481.33+OII\\
    \bottomrule
  \end{tabular}
\end{table}\clearpage

\begin{table}[!t]\centering
  \setlength{\tabnotewidth}{0.9\textwidth}
  \tablecols{4}
  \setlength{\tabcolsep}{\tabcolsep}

\caption{ Mean radial velocities and equivalent widths in October 2001}
\begin{tabular}{lrrr}
\toprule
Line ID & RV &GFWHM& EW (\AA) \\
  &       km s$^{-1}$ &km s$^{-1}$&\AA  \\
(1)&(2)&(3) & (4) \\
\midrule
 H I       &  &  &    \\
  4340.47 &     -26  &  250 &0.9       \\
  4861.33 &     -39  &  200 &0.9     \\
 HeI         &  &  &    \\
  4387.93  &    -20 & 220   &0.52   \\
  4471.58  &    -30 & 200   &1.06   \\
  4713.26  &    -17 & 150   &0.38   \\
  4921.93  &    -14 & 140   &0.63   \\
  5015.68  &    -35 & 130   &0.33   \\
  5875.80  &    -43 & 160   &0.73    \\
  6678.15  &     -  & 150   &0.85     \\
 C III        &  &  &    \\ 
  4388.02   &  -36  & 166   & 0.51           \\
 N II          &  &  &    \\
  5676.02    &    -12& 100 & 0.1:                 \\
  5679.56   &     -10& 100:& 0.23                \\
 O II       &     &  &    \\ 
  4661.64   & +3  & 50 & 0.16          \\
  5508.11   & -14: &120&0.1:            \\
 Si II      &    &  &    \\
  5219.37   & -24 &130 &0.1:            \\
 Si III         &  &  &    \\ 
  4552.62   &  -5 &128 &0.40            \\
  4567.83   &  -8 &120 &0.45                    \\
  5739.73   &  -9 &135 & 0.30                   \\ 
 Mg II      &     &    &                        \\
  4481.23  &  -18:&140:& 0.2          \\
    \bottomrule
  \end{tabular}
\end{table}\clearpage

\onecolumn

\begin{figure}
\includegraphics[width=\columnwidth]{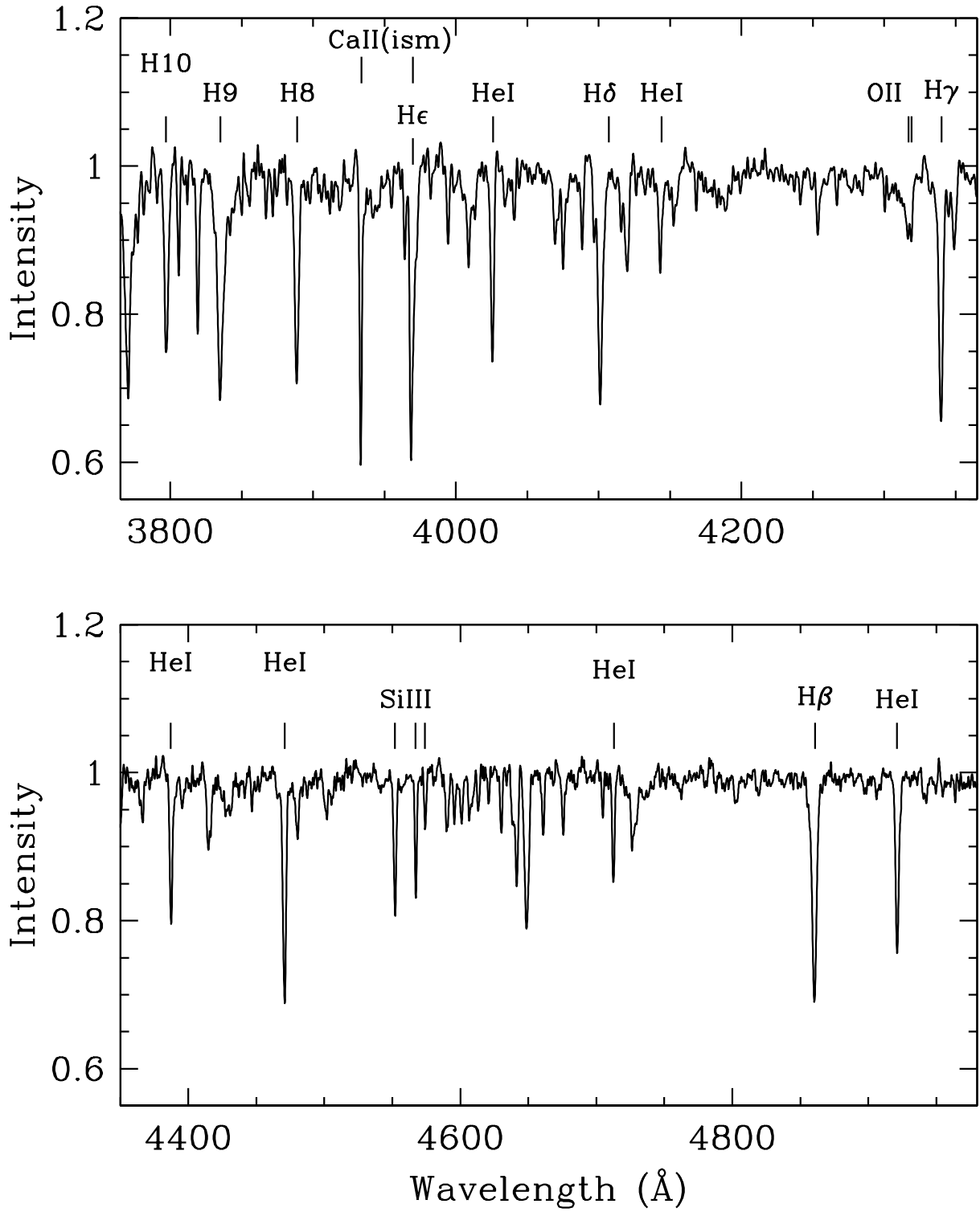}
\caption{A portion of the normalized spectrum of LS {\rm I}+65 010.
\label{fig1}}
\end{figure}
\clearpage

\begin{figure}
\includegraphics[width=\columnwidth]{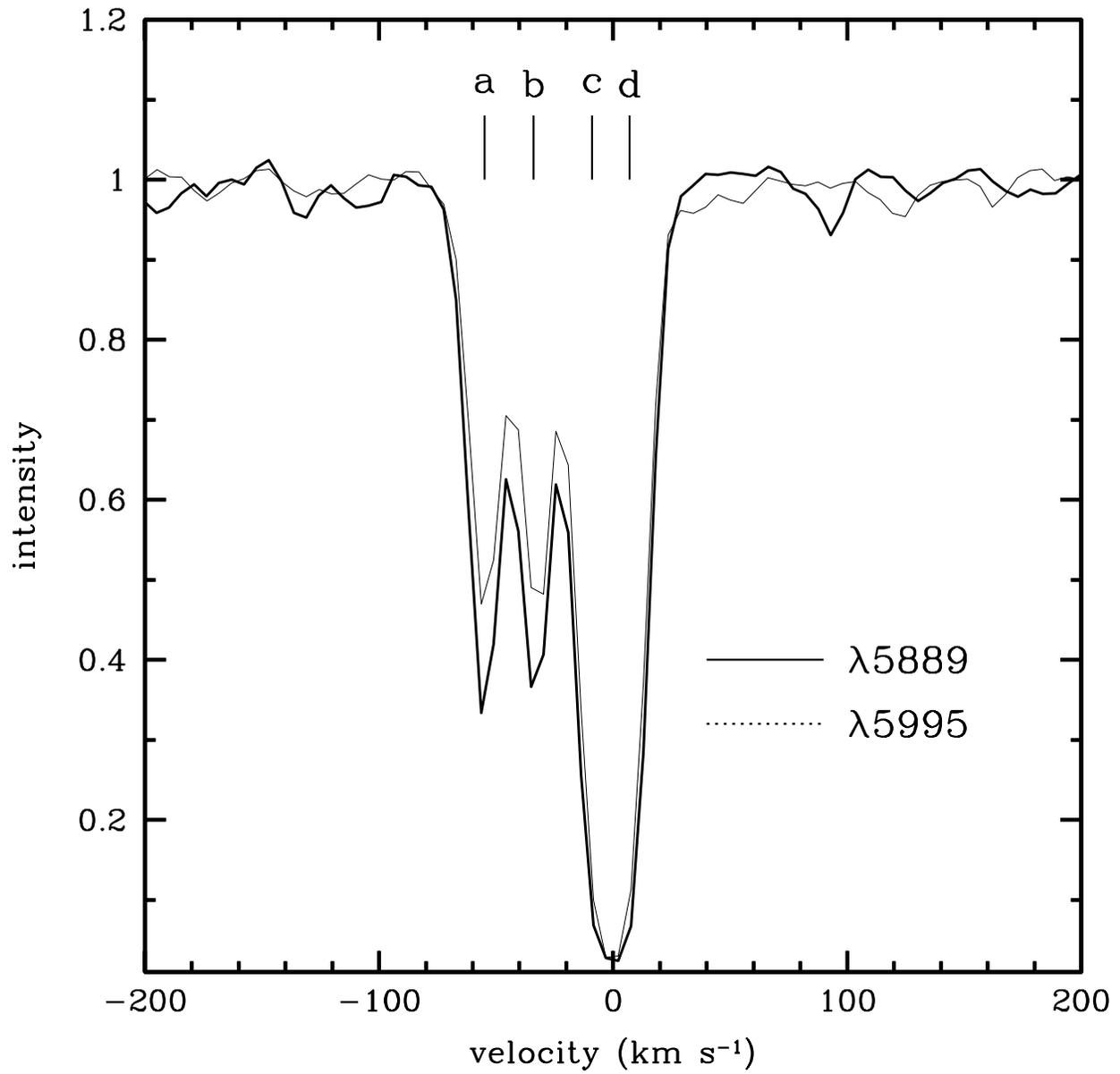}
\caption{Plot of the \ion{Na}{1} $\lambda$5889 (dark) and $\lambda$5895
(light) line profiles observed in the coud\'e spectra of LS {\rm I}+65 010.
The tick marks indicate the central position of the four Gaussian profiles
that were fit to the blend.  
\label{fig2}}
\end{figure}
\clearpage

\begin{figure}
\includegraphics[width=\columnwidth]{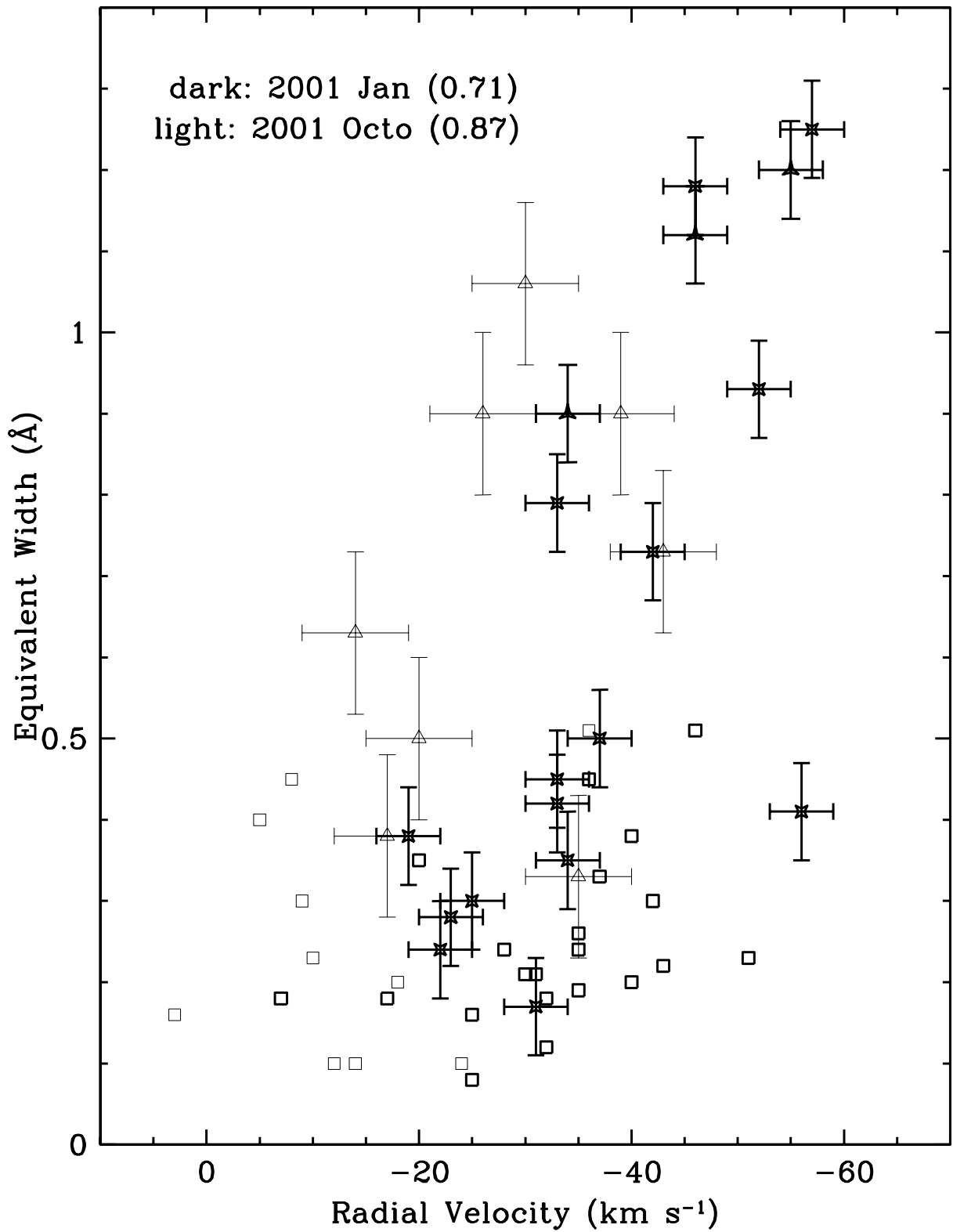}
\caption{Equivalent width of the photospheric absorption lines plotted against 
the
velocity of the line, for the data of January 2001 (dark) and October 2001 
(light).
Triangles in the January data represent H-lines while crosses represent
He I lines.  The open squares represent lines from heavier elements (see
Tables 8 and 9).  
\label{fig3}}
\end{figure}
\clearpage

\begin{figure}
\includegraphics[width=\columnwidth]{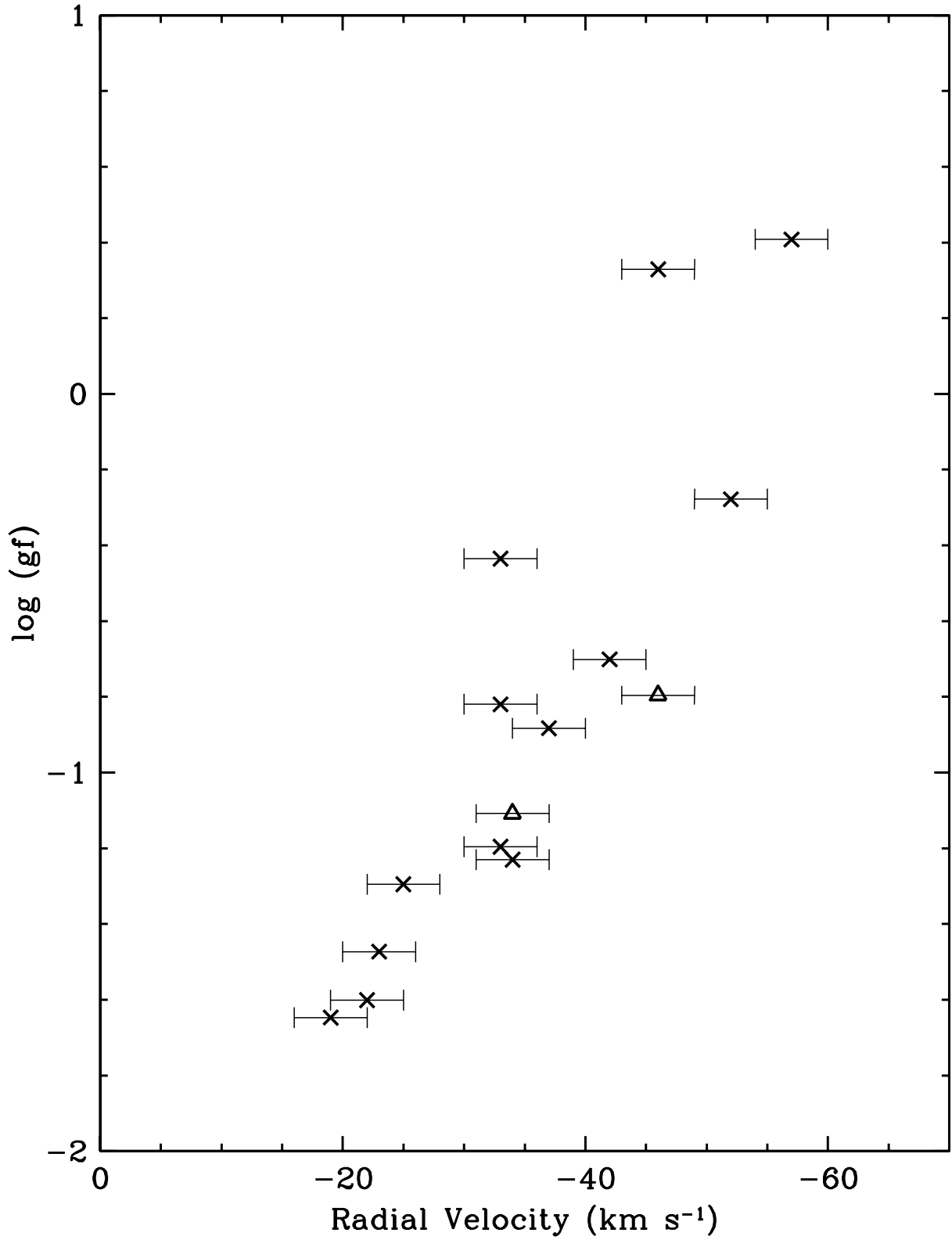}
\caption{Correlation of log(gf) {\it vs.} RV for the H and \ion{He}{1} lines 
in
the 2001 January data.
\label{fig4}}
\end{figure}
\clearpage

\begin{figure}
\includegraphics[width=\columnwidth]{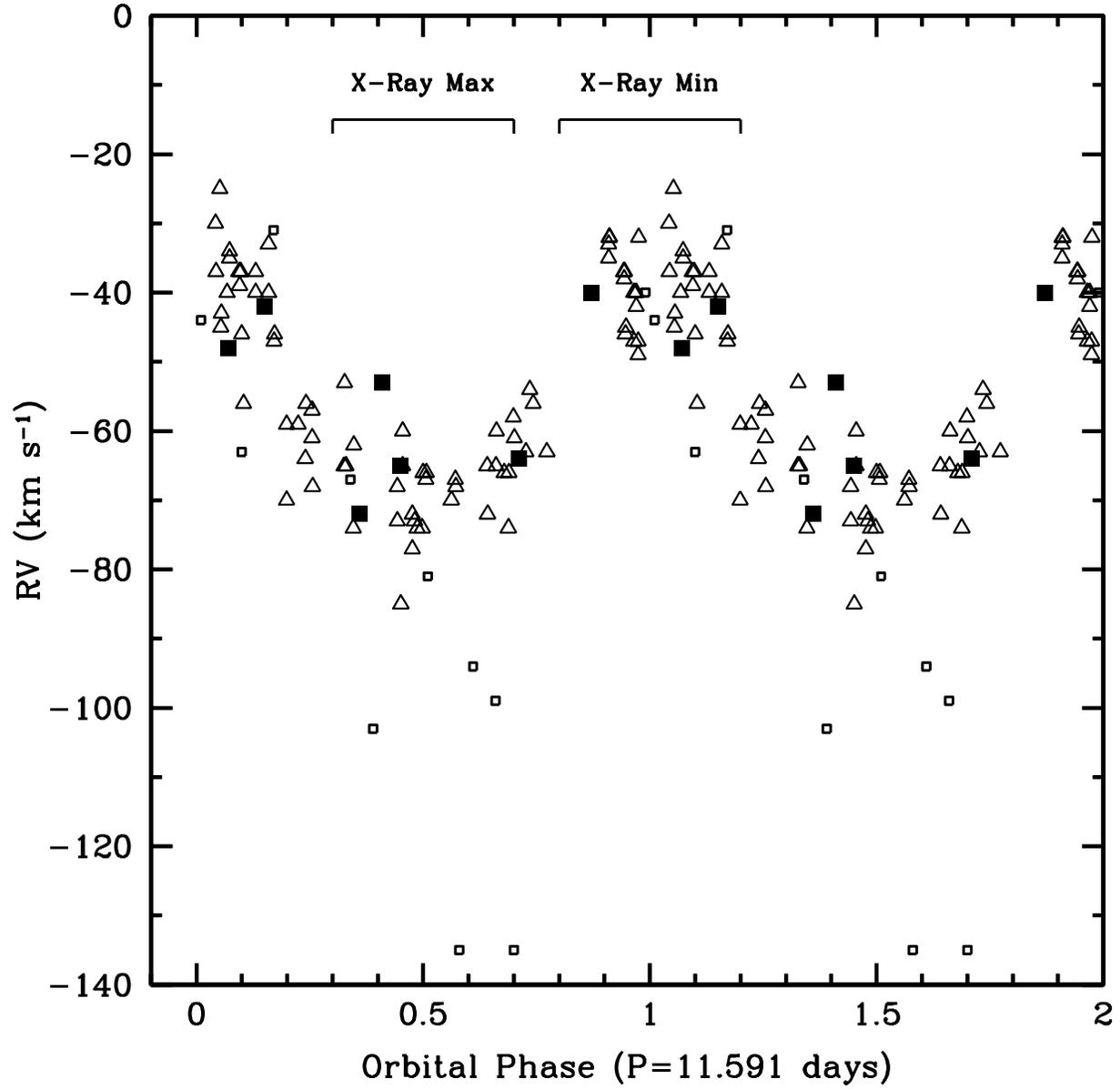}
\caption{Radial velocities folded in phase with the 11.591 day orbital 
period.  Open triangles are the data of Crampton et al.,
open squares are the data published by Reig et al. for \ion{He}{1} 6678,
and filled squares are our Vel$^{One}$ data for He I 5875 \AA. Indicated are 
the approximate
phase intervals during which the RXTE X-ray counts were maximum and minimum
(from Fig. 7 of Corbet et al. 1999). The neutron star is ``in front" at phase
0.25.
\label{fig5}}
\end{figure}
\clearpage

\begin{figure}
\includegraphics[width=\columnwidth]{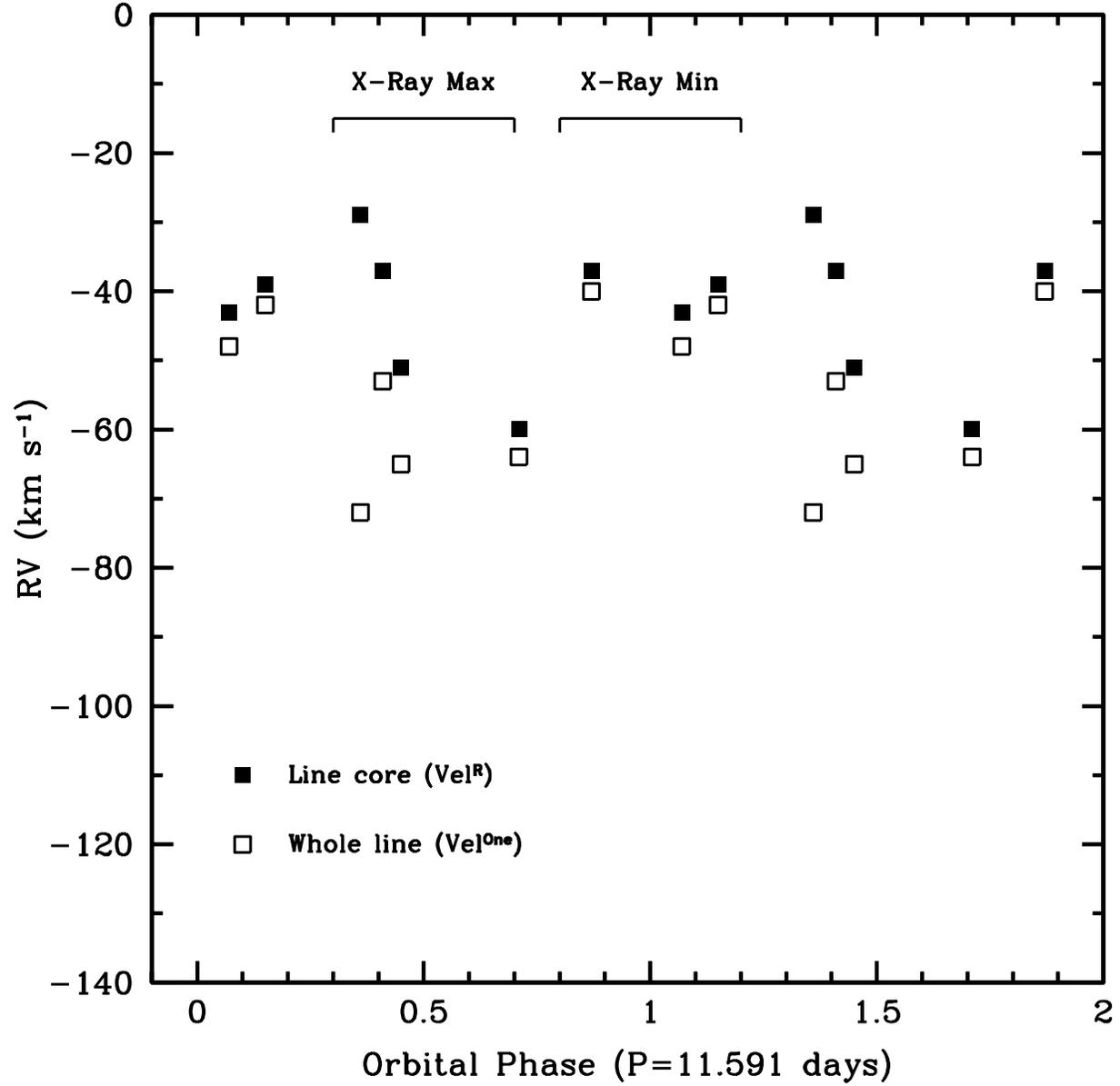}
\caption{Radial velocities of He I 5875 \AA\ folded in phase with the 11.591 
day 
orbital period: results obtained by measuring the whole line, using a single
Gaussian fit (Vel$^{One}$; open squares) compared with the measurements of 
only the line
core (Vel$^R$; filled-in squares).  The two measurements have their largest 
discrepancy 
at phases when the X-ray flux is maximum.
\label{fig6}}
\end{figure}
\clearpage

\begin{figure}
\includegraphics[width=\columnwidth]{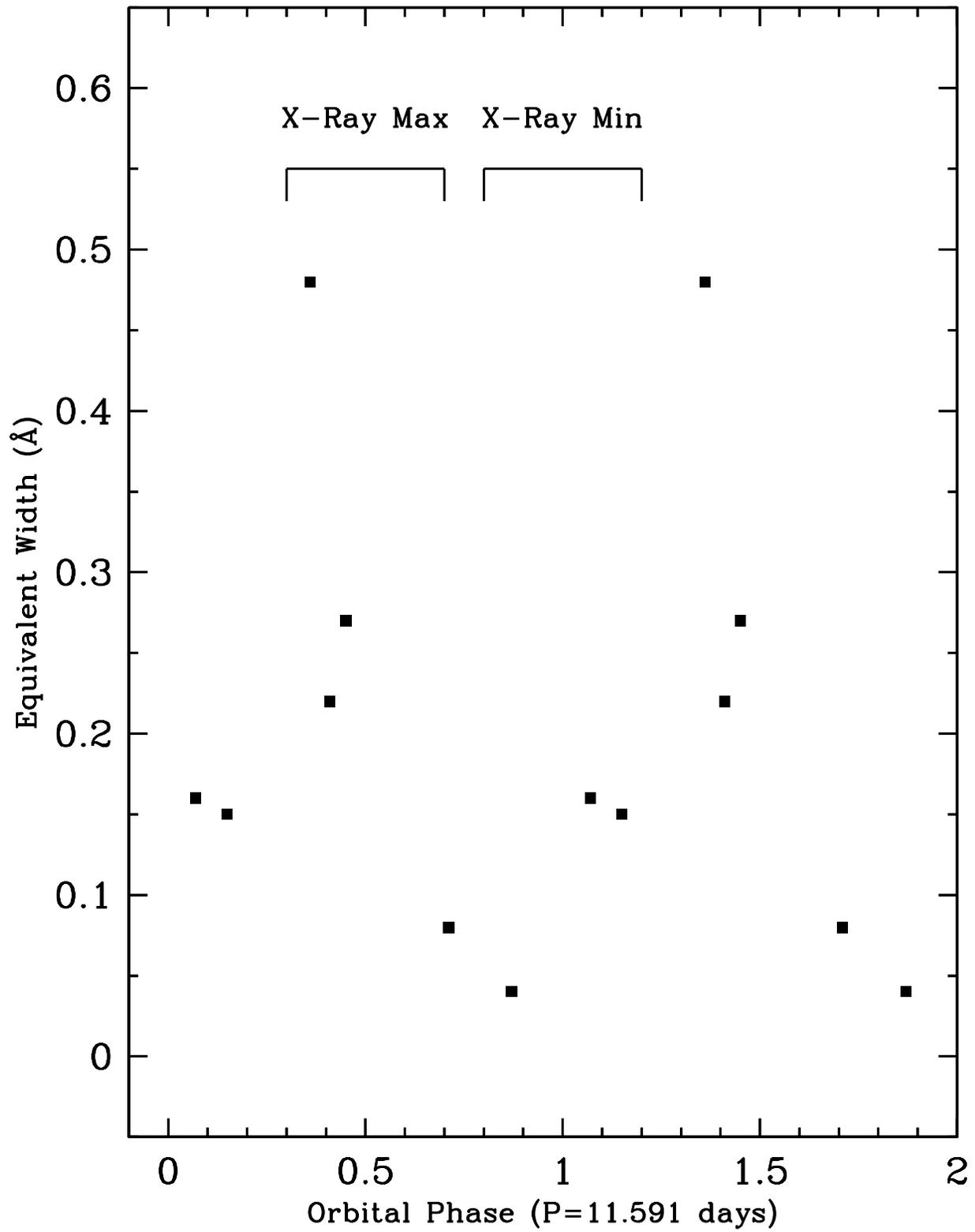}
\caption{Equivalent width of the ``blue" Gaussian that was fit to the He I 
5875 line
profile, as a function of the 11.591 day orbital phase.
\label{fig7}}
\end{figure}
\clearpage

\begin{figure}
\includegraphics[width=\columnwidth]{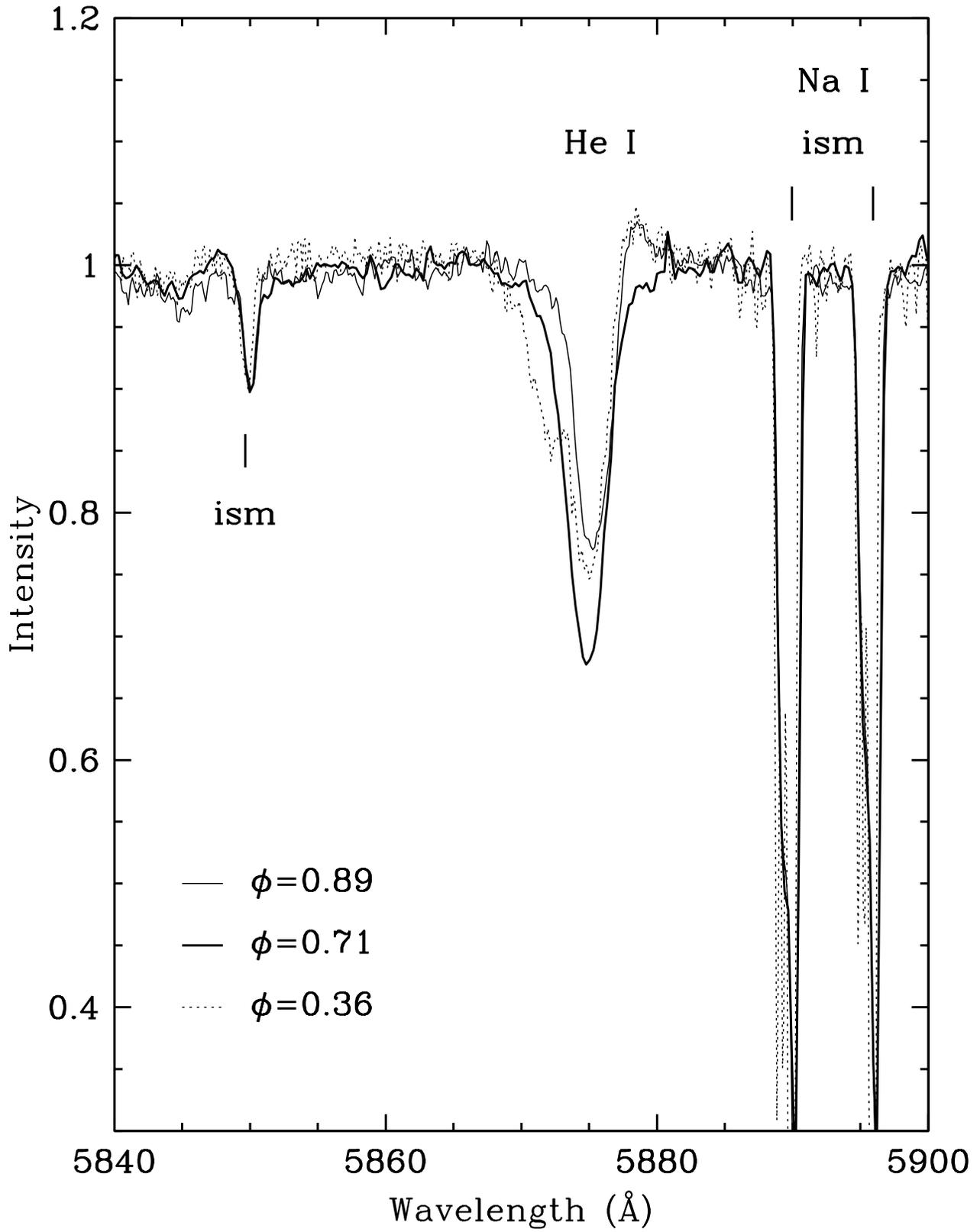}
\caption{Comparison of the He I 5875 line  profiles at different
orbital phases, as indicated. At orbital phase 0.71 (dark tracing), the
collapsed object is on the far side of the B1-star, and the photospheric
absorption line is deeper and more symmetric than at other phases.
\label{fig8}}
\end{figure}
\clearpage

\begin{figure}
\includegraphics[width=\columnwidth]{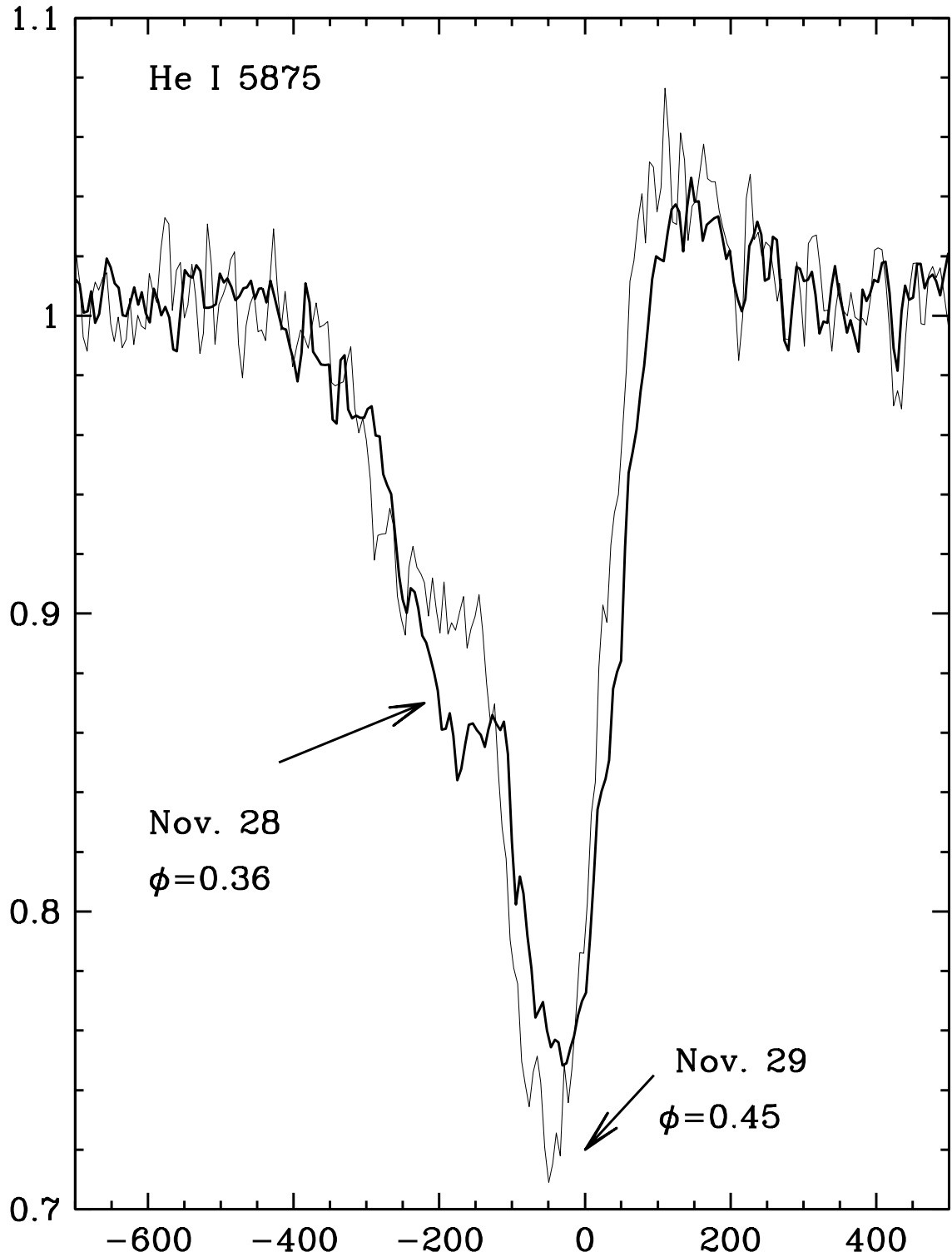}
\caption{Comparison of the He I 5875 profiles in the 1995 coude
data.  Note that the core of the line profile  of 29 Nov. (dotted tracing) is
displaced blueward with respect to the line profile of the previous night by 
16 km s$^{-1}$.
\label{fig10}}
\end{figure}
\clearpage

\begin{figure}
\includegraphics[width=\columnwidth]{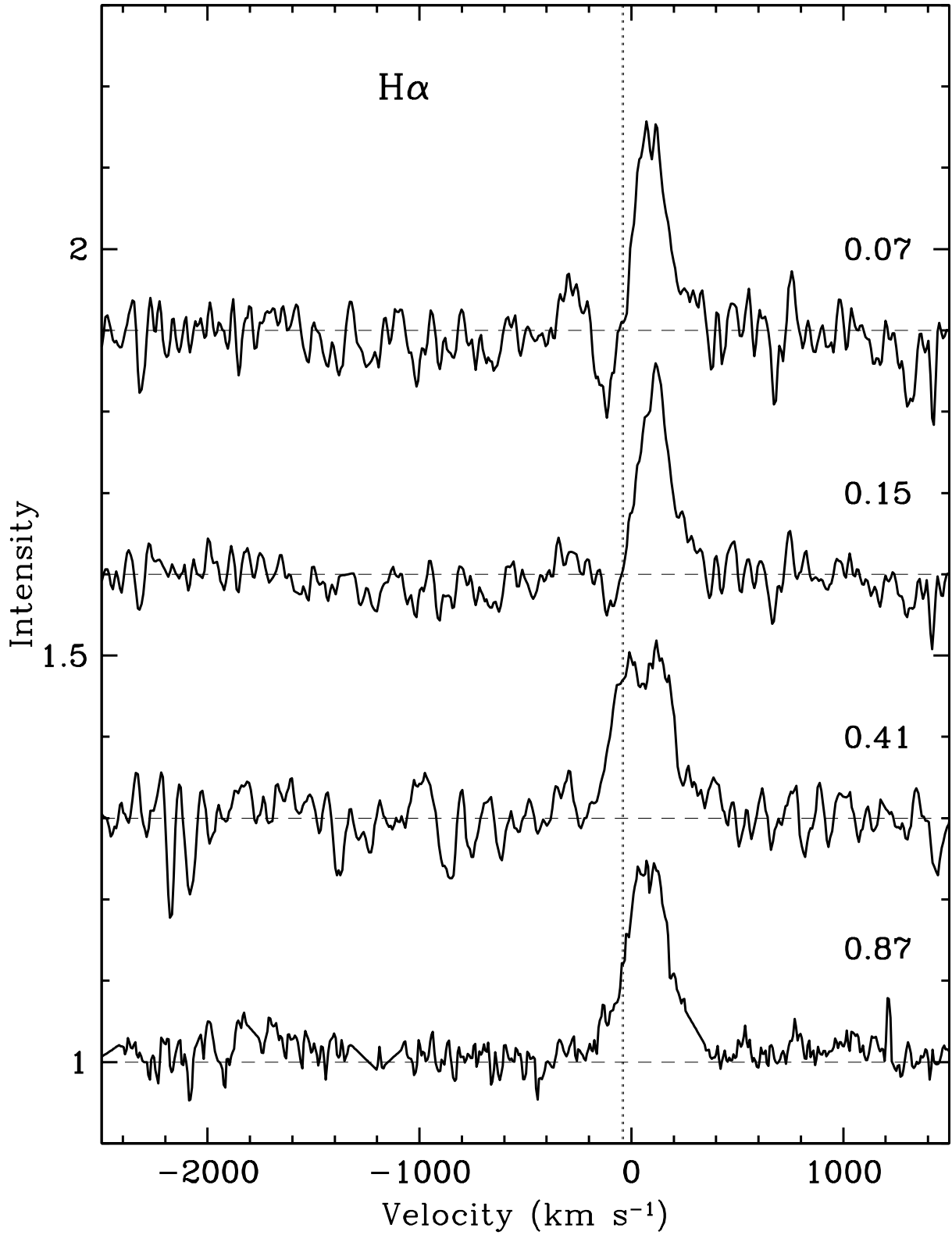}
\caption{Comparison of the H$\alpha$ line  profiles at different
orbital phases, as indicated.
\label{fig11}}
\end{figure}
\clearpage

\begin{figure}
\includegraphics[width=\columnwidth]{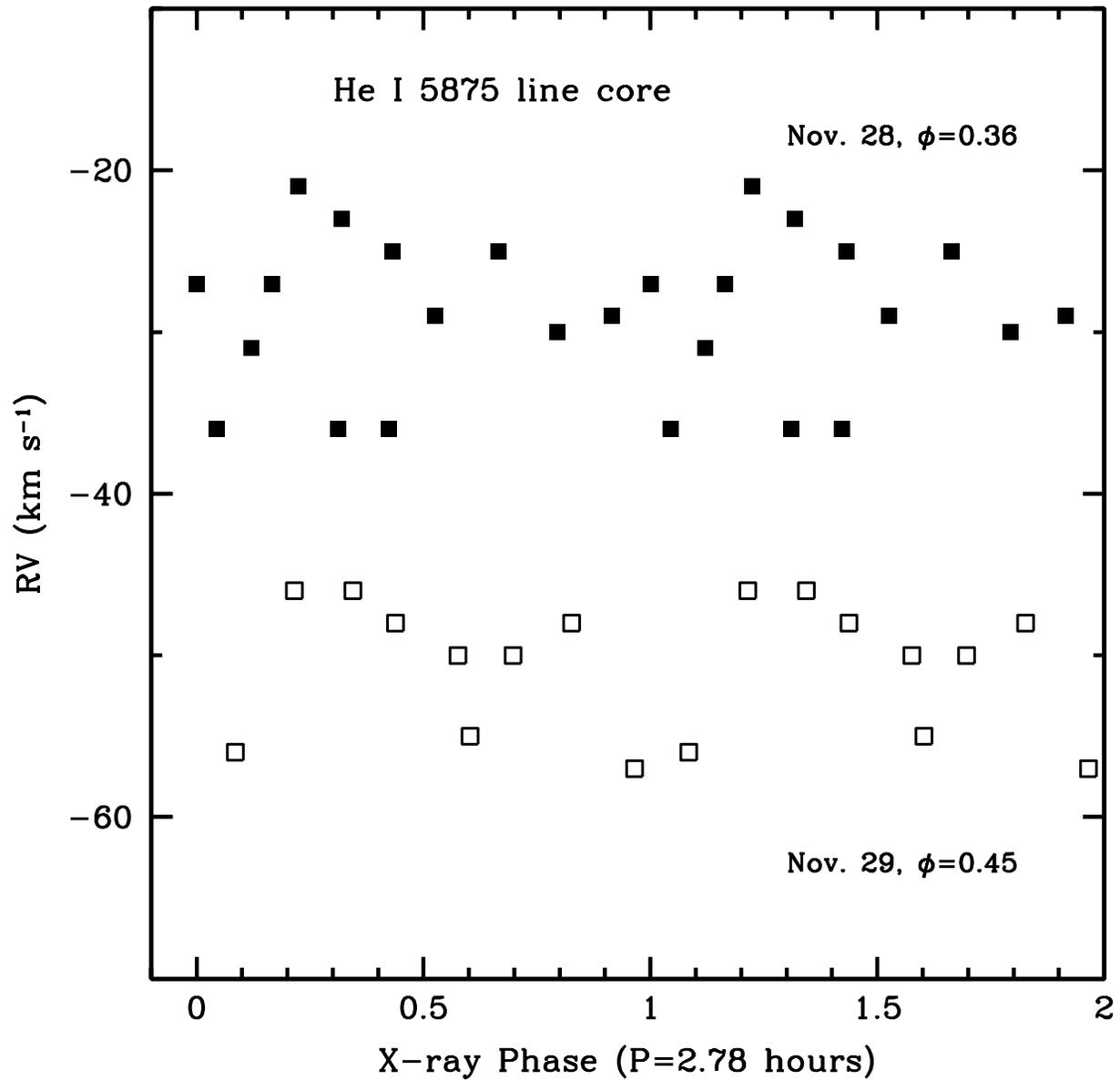}
\caption{Radial velocity of the ``red" Gaussian (i.e., fit to the line core)
of the He I 5875 absorption line in the 1995 data, plotted against 2.78 hour 
X-ray
phase.  Filled squares are Nov. 28 data and open squares are Nov. 29 data. 
Although variability is present on the short-term, it does not appear to 
follow
the 2.78 hour X-ray flaring period.
\label{fig12}}
\end{figure}

\begin{figure}
\includegraphics[width=\columnwidth]{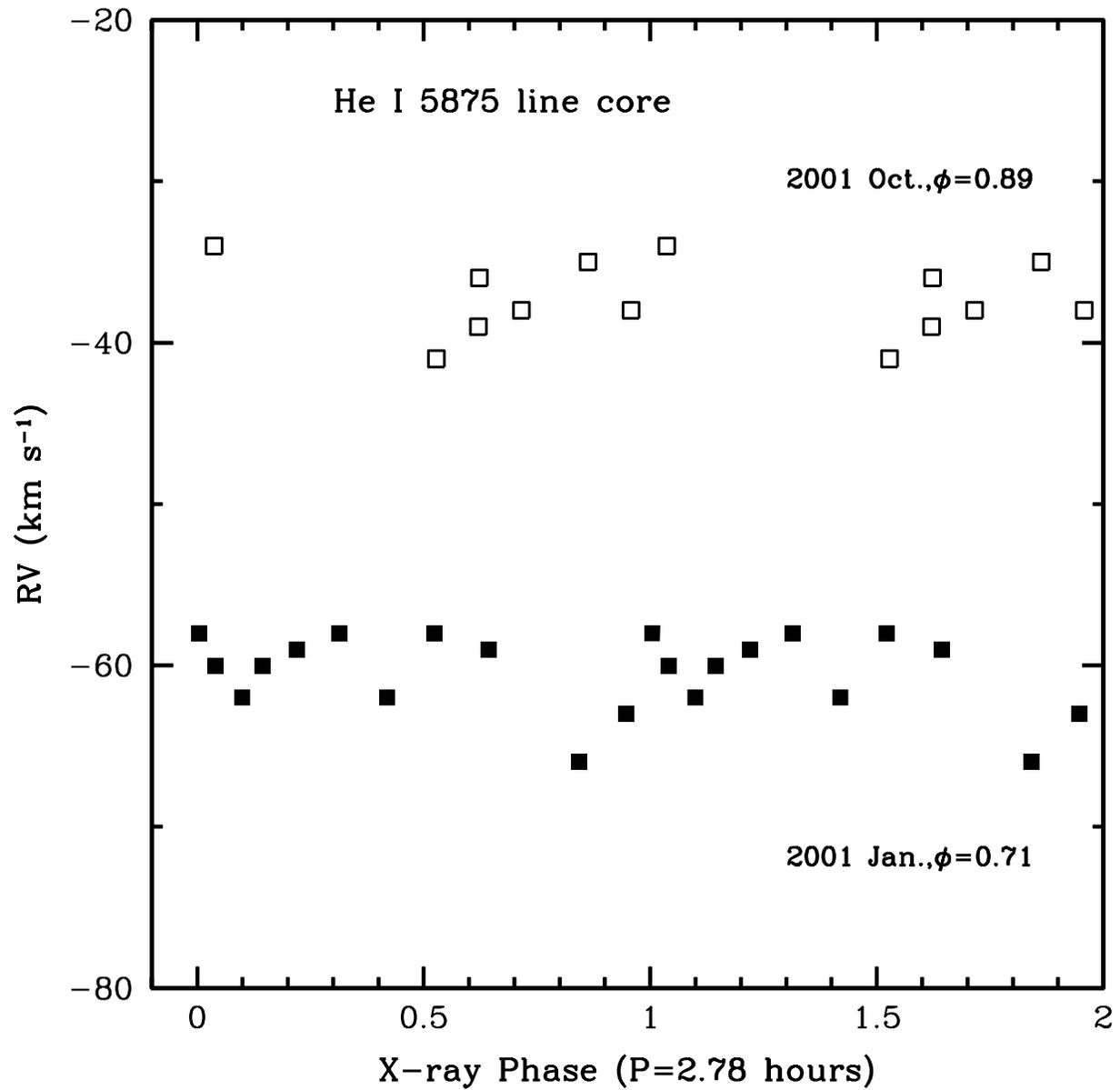}
\caption{Radial velocity of the ``red" Gaussian (i.e., fit to the line core)
of the He I 5875 absorption line in the 2001 data, plotted against 2.78 hour 
X-ray
No variability greater than the $\pm$5 km s$^{-1}$ uncertainties is present.
phase. 
\label{fig13}}
\end{figure}
\clearpage

\begin{figure}
\includegraphics[width=\columnwidth]{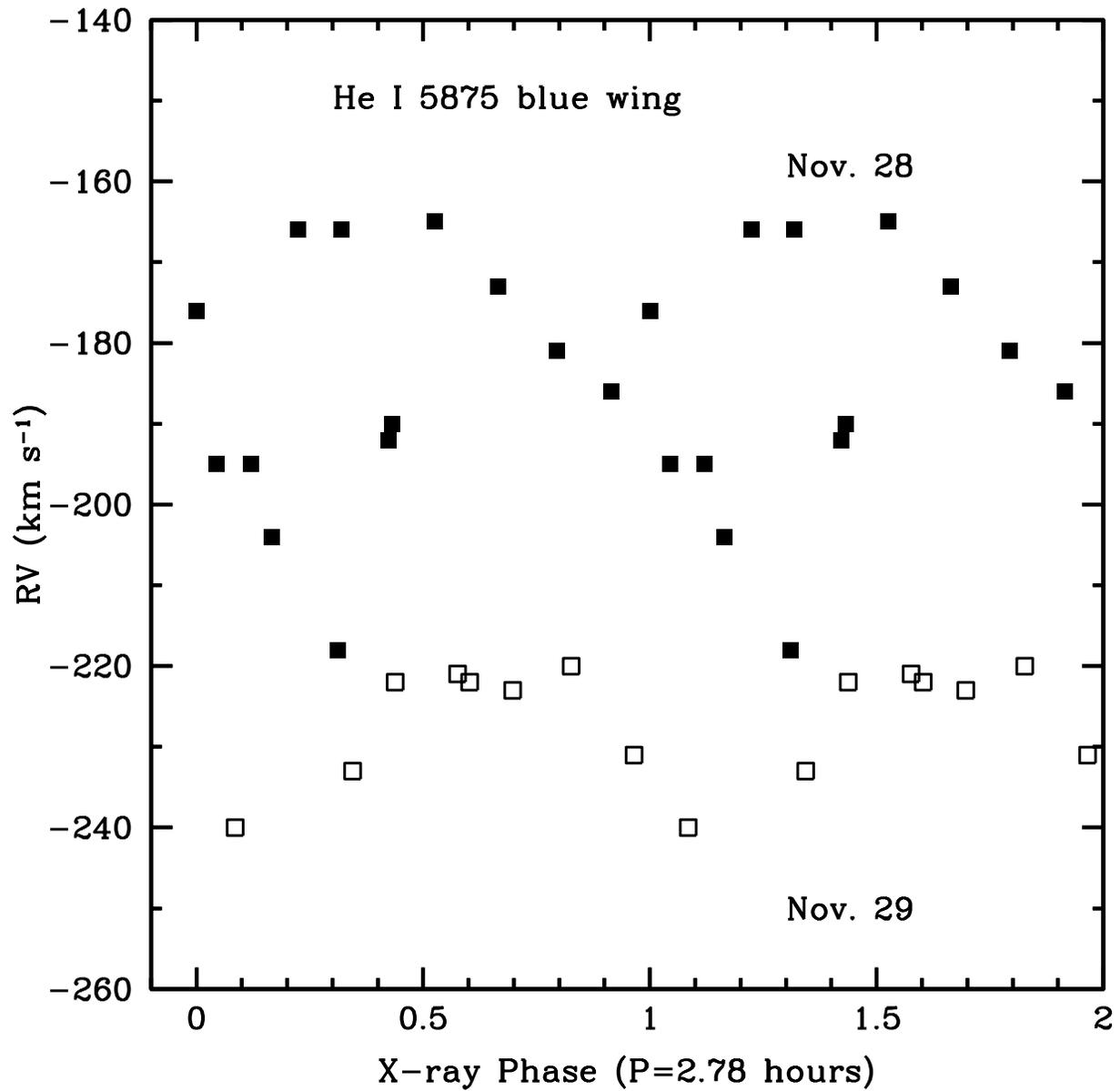}
\caption{Radial velocity of the ``blue" Gaussian (i.e.,fit to the blue line 
wing)
of the He I 5875 absorption line in the 1995 data, plotted against 2.78 hour 
X-ray
phase.  Filled squares are Nov. 28 data and open squares are Nov. 29 data. 
\label{fig14}}
\end{figure}
\clearpage

\end{document}